\documentclass[twocolumn,twoside]{article}
\usepackage[utf8]{inputenc}
\pdfoutput=1
\usepackage[english]{babel}

\usepackage[letterpaper,top=2cm,bottom=2cm,left=3cm,right=3cm,marginparwidth=1.75cm]{geometry}

\usepackage{amsmath}
\usepackage{graphicx}
\usepackage{authblk}
\usepackage{abstract}
\usepackage[colorlinks=true, allcolors=blue]{hyperref}

\title{Hatemongers ride on echo chambers to escalate hate speech diffusion}
\author[a]{Vasu Goel}
\author[a]{Dhruv Sahnan}
\author[b]{Subhabrata Dutta}
\author[c]{Anil Bandhakavi}
\author[b]{Tanmoy Chakraborty}

\date{}
\affil[a]{Department of Computer Science \& Engineering, IIIT Delhi, India 110020}
\affil[b]{Department of Electrical Engineering, IIT Delhi, India 110016}
\affil[c]{Logically, EC1A 9HAUK, UK}
\begin{document}
\twocolumn[
    \begin{@twocolumnfalse}
    \maketitle
        \begin{center}
            \begin{abstract}
Recent years have witnessed a swelling rise of hateful and abusive content over online social networks. While detection and moderation of hate speech have been the early go-to countermeasures, the solution requires a deeper exploration of the dynamics of hate generation and propagation. We analyze more than 32 million posts from over 6.8 million users across three popular online social networks to investigate the interrelations between hateful behavior, information dissemination, and polarised organization mediated by echo chambers. We find that hatemongers play a more crucial role in governing the spread of information compared to singled-out hateful content. This observation holds for both the growth of information cascades as well as the conglomeration of hateful actors. Dissection of the core-wise distribution of these networks points towards the fact that hateful users acquire a more well-connected position in the social network and often flock together to build up information cascades. We observe that this cohesion is far from mere organized behavior; instead, in these networks, hatemongers dominate the echo chambers -- groups of users actively align themselves to specific ideological positions. The observed dominance of hateful users to inflate information cascades is primarily via user interactions amplified within these echo chambers. {We conclude our study with a cautionary note that popularity-based recommendation of content is susceptible to be exploited by hatemongers given their potential to escalate content popularity via echo-chambered interactions.}
            \end{abstract}
         \end{center}
     \end{@twocolumnfalse}
]
% \section*{Siginificance statement}
% This work investigates the spread of online hate speech through the lens of information propagation, user engagement, and polarised consumption/production of information via echo chambers. Our findings establish that hatemongers are way more potent to control the information dissemination over online social networks, even with apparently non-hateful introductions, compared to singled-out hateful posts. This is largely due to such users acquiring a well-connected position on these networks through cohesive engagement with other hateful actors. Unlike merely organized activities, we find that this cohesion among hatemongers is realized via echo chamber formation. We show that the amplified potential of hatemongers to popularize content is controlled via their echo chamber memberships.

\section*{Introduction}
The early upheaval of online social networks like Facebook, Twitter, Reddit, etc. to revolutionize the mode of communication and day-to-day information consumption has started to saturate. From the standpoint of end-users as information consumers, their presence in everyday life is now ubiquitous \cite{waheed2017investigation}. While this has significantly increased worldwide connectivity and information production/consumption, it is {\em no free lunch}. In the past decade, the world has observed a staggering rise in polarization \cite{simchon2022troll}, abusive content, and misinformation dominating the online social space \cite{doi:10.1073/pnas.1517441113,doi:10.1073/pnas.2022819118,doi:10.1073/pnas.2013464118}. A recent survey has reported that around $41\%$ of the US population have been on the receiving end of some hateful behavior at least once in their life \cite{league2020online}. Furthering the peril, online hate speech has transcended the virtual to sprinkle vitriol into the real \cite{AWAN20161, real-hate, hate-myanmar}. 

The research community has engaged in this arena with increasing efforts as well. Multiple meta-analyses have suggested a superlinear growth in research related to hate speech in recent years \cite{meta-hate-1, meta-hate-2,Cinelli2021_hate}. Most of these studies seek to {\em identify} hate speech; some explore the {\em dynamics} as well \cite{Velasquez2021-covid-hate, Uyheng2021, 9458789, Johnson2019-hate-network}. The latter is particularly of interest for combating the spread of hate speech since only content moderation via flagging, banning, or deleting posts may not be enough in this context \cite{hate-regulation, hate-content-moderation, Johnson2019-hate-network} (it may often incur threats to the democratic principles~\cite{hate-undemocracy}). It is unanimously agreed that certain malicious groups take advantage of the apparent anonymity on these platforms to create and propagate hateful content \cite{real-hate-white-supre, winter2019online, hate-spreader-theory}. However, it is unlikely that a handful of malevolent actors could dictate the large-scale characteristics of such platforms; the inner workings of these platforms \cite{algorithmic-amplification}, reinforced by the real-world social processes \cite{youngblood2020extremist}, should be investigated for how they prepare the breeding ground for online hatemongering. Two separate earlier findings in this context prepare the foundation of our current study. Firstly, the diffusion of information over a platform, whether mediated by hateful actors (users) or via hateful content (posts), exhibits different characteristics compared to their non-hateful counterparts \cite{Cinelli2021_hate, 9458789}. Secondly, in the social science community, it has been conjectured that hateful and extremist behavior might be linked with the formation of {\em echo chambers} \cite{misogynist-manosphere, taub2018social} -- groups of users who share a strong opinion align themselves in the interaction network in such a way that they are exposed to content correlated to their chosen ideology.

Our study spans over three popular social media platforms: Reddit, Twitter, and Gab --- the first one is a strongly moderated discussion forum, whereas the latter two are microblogging sites with partial to no moderation (Twitter is not as carefully moderated as Reddit, whereas Gab is an unmoderated platform to promote ``freedom of speech'') \cite{Artime2020,jhaver_mod}. Therefore, the chosen social media platforms are expected to cover different aspects of online discussion. We collect and analyze a total of $6.8$ million users across these three platforms, covering over $32$ million posts and over $0.1$ million information cascades. 

We start by analyzing information cascades characterized by the hatefulness of the source content as well as of the users introducing them to the platforms. An observation common across all three platforms is that {\em hate attracts hate} --- hateful posts/users cluster around a source post/user more if the latter is hateful. However, there is a remarkable distinction in the importance of the type of content vs. the type of the user in terms of procuring further engagement. We observe that {\em hateful users are more prepollent than hateful content}. Content posted by a highly-hateful user is likely to attract more engagement compared to the same posted by a low-hateful user; even non-hateful posts from high-hateful users tend to catalyze larger cascades, compared to posts from low-hateful users. Upon analyzing the hate characteristics of these cascades, we find that the proportion of hateful participation in the cascade is also larger when the source user is hateful compared to when not. This observation further strengthens the claim that mere content moderation is not enough to combat hate speech.

Further investigations unravel the underlying user interaction dynamics, leading to the observed information dissemination characteristics. We notice increasing user hatefulness as we move towards the network cores. This, along with the observed affinity of hateful users to cluster around hateful users (even when they do not post anything hateful specifically) across all the platforms, drives our focus towards investigating the formation of echo chambers and their relation to hatefulness \cite{gillani2018me,quattrociocchi2017inside,Garimella,doi:10.1073/pnas.2023301118}. To this end, we propose a novel method of echo chamber discovery in online social networks, developed upon the operational formalism of echo chambers defined in \cite{doi:10.1073/pnas.1517441113,Sasahara2019OnTI,Choi2020}. Unlike relying on indirect cues of opinion affinity used by earlier works (e.g., URLs), we directly utilize the content posted by users to define `opinion ecology'. We then define an `echo chamber' as a set of users with highly shared ideology (homophily) and selective exposure to an opinion \cite{doi:10.1073/pnas.2023301118,Choi2020}.

Analyses of the echo chambers discovered using this method empirically validate the hitherto conjecture that {\em hateful behavior over online social networks does intensify through echo chamber formation}. The boost in the volume for cascades originating from highly-hateful users is shown to be directly attributed to the user's affiliation to echo chambers. Furthermore, the cascade participation is strongly biased towards users in echo chambers when the source user is also a member. Finally, we assign a {\em homogeneity score} to an echo chamber based on the degree of mixing of high-hateful vs. low-hateful users as the constituents ---
a highly-pure echo chamber would primarily consist of either high-hateful or non-hateful users.
A strong positive correlation between hatefulness and homogeneity is observed across all three social networks; pure echo chambers are predominantly hateful. However, the homogeneity distribution is skewed for different networks -- while Gab exclusively contains pure and predominantly hateful echo chambers, Reddit shows a wider spectrum. We conclude this as further evidence of the interrelation between intense polarization and the spread of hate speech over social networking platforms. {Since features like {\em Top posts}, {\em Hot topics}, {\em Trending now}, etc. provided by several platforms rely on ranking posts/topics based on the user engagement they receive and draw the attention of other users to them, echo chamber-driven amplification of hatemonger influence can be a critical factor to keep in mind while designing countermeasures.}

\section*{Characterizing Hate and Echo Chambers}

\subsection*{Social Networks Investigated}

We investigate three popular online interaction platforms: Reddit, Twitter, and Gab. The interaction scopes defined for the users on these platforms are very distinct. Reddit is primarily designed as a discussion platform with a very limited scope of sharing already posted content. Also, user interactions on Reddit are governed by numerous user-defined communities {\em aka} {\em subreddits} that predefine the broad topics of discussions with varying degrees of decentralized moderation (i.e., each subreddit has its own set of moderation rules, moderator activity, etc.). Twitter and Gab, on the other hand, are predominantly used as information-sharing platforms via posting and resharing; while there are scopes to reply to a certain post back and forth to construct `discussions', they are rarer as well as smaller compared to resharing-based cascades. Furthermore, Twitter enforces some degree of content moderation in a centralized manner that has been reformulated and reimplemented multiple times in the past; very often, exclusively hateful tweets from users get deleted early on~\cite{tweet-removal}.  Gab, on the other hand, is maintained as an alter ego of Twitter with absolute freedom of speech \cite{Zannettou2018WhatIG}.

Reddit is composed of submissions and comments. We connect users based on their commenting behavior on different submissions and comments. We follow the method used in \cite{doi:10.1073/pnas.2023301118} to create our network. A directed link from user $x$ to user $y$ exists if $y$ has replied to a comment or submission from $x$. 
We collect our data from various controversial subreddits for the year 2019 (e.g., controversy, men's rights, environment, etc.), covering over $0.8$ million posts from over $97$ thousand users with more than $0.4$ million unique user interactions.

Twitter is made up of posts and retweets. We connect users based on their retweeting behavior on different tweets. We follow the method used by \cite{doi:10.1073/pnas.2023301118} to construct our network \textcolor{black}{with tweets within April to June, 2019}. A directed link from user $x$ to user $y$ exists if $y$ has retweeted or quote-tweeted a tweet from $x$. Altogether, it covers over $15$ million unique user interactions among $67$ million users. A total of $315$ million tweets are analyzed.

Gab is made up of submissions and comments. Similar to Twitter, users can follow each other. For our analysis, we place a directed link from user $x$ to user $y$ if $y$ has reposted or quoted $x$'s post. The repost and quote work in a similar fashion as retweet and quote tweet, respectively. The collected user network consists of $29$ thousand users with over $0.1$ million unique interactions through $0.3$ million posts \textcolor{black}{appeared within October, 2020 to September, 2021}.

Readers may refer to the {\color{blue} Materials and Methods} and {\color{blue}{\em SI Appendix}, Section 1} for more details about the datasets. 

\subsection*{Identification of Hateful Content and Users}

Several studies on large-scale hate speech detection use a predefined set of lexicons to identify a piece of content as hateful \cite{davidson2017automated,qian-etal-2021-lifelong}. However, the applicability of such lexicons can become very limited once the topic of discussion shifts. We refrain from defining hate speech on our own. Instead, we rely on existing hate speech detectors. \textcolor{black}{Identification of hate speech strongly depends on the context under consideration, e.g., type of the event being discussed, time-frame, target of the hate speech being directed towards, discussion forum, etc. 
 To circumvent this, we use an ensemble of multiple classifiers trained using different types of hatefulness datasets.} We label the degree of hatefulness of each post based on three different state-of-the-art hate speech detectors -- Davidson \cite{davidson2017automated}, Waseem \cite{waseem2016hateful}, and Founta \cite{Fountana}. A post is tagged as {\em non-hateful} when {\em all} the detectors decide it to be non-hateful. If two or more classifiers find them hateful, then we categorize it to be {\em highly-hateful}, otherwise {\em medium-hateful}. \textcolor{black}{Further discussion on the detection of hate speech is provided in {\color{blue}{\em SI Appendix},  Section 1.}}

We further classify each user into one of the three buckets based on the hatefulness as suggested in \cite{10.1145/3292522.3326034} --
%\begin{enumerate}
    \textbf{high: } if the user posted five or more hateful posts (medium and/or high),  \textbf{medium: } if the user posted two or more hateful posts,  \textbf{low: } if the user posted one or no hateful post.

\begin{figure}
\centering
\includegraphics[width=\textwidth]{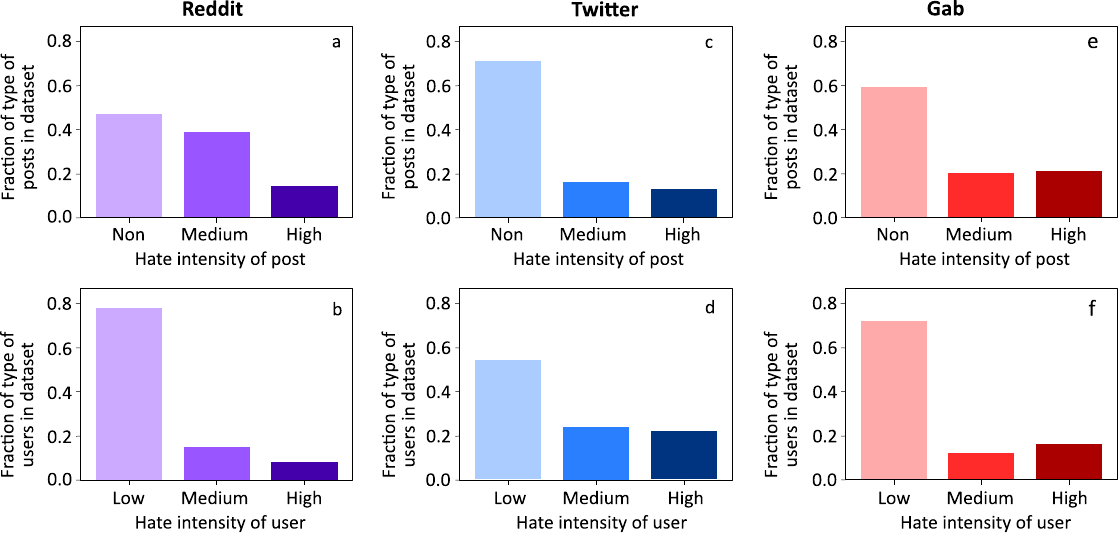}
\caption{Fraction of type of posts (subplots a., c., e.) and users (subplots b., d., f.) characterized based on their hatefulness (low, medium, and high).}
\label{SI:dataset_comp_hatefulness}
\end{figure}

We illustrate the percentage distribution of hateful posts and users as classified by our method for all three datasets in {\color{blue} {\em SI Appendix}, Figure \ref{SI:dataset_comp_hatefulness}}.

\begin{figure*}[!t]
    \centering
  \includegraphics[width=\textwidth]{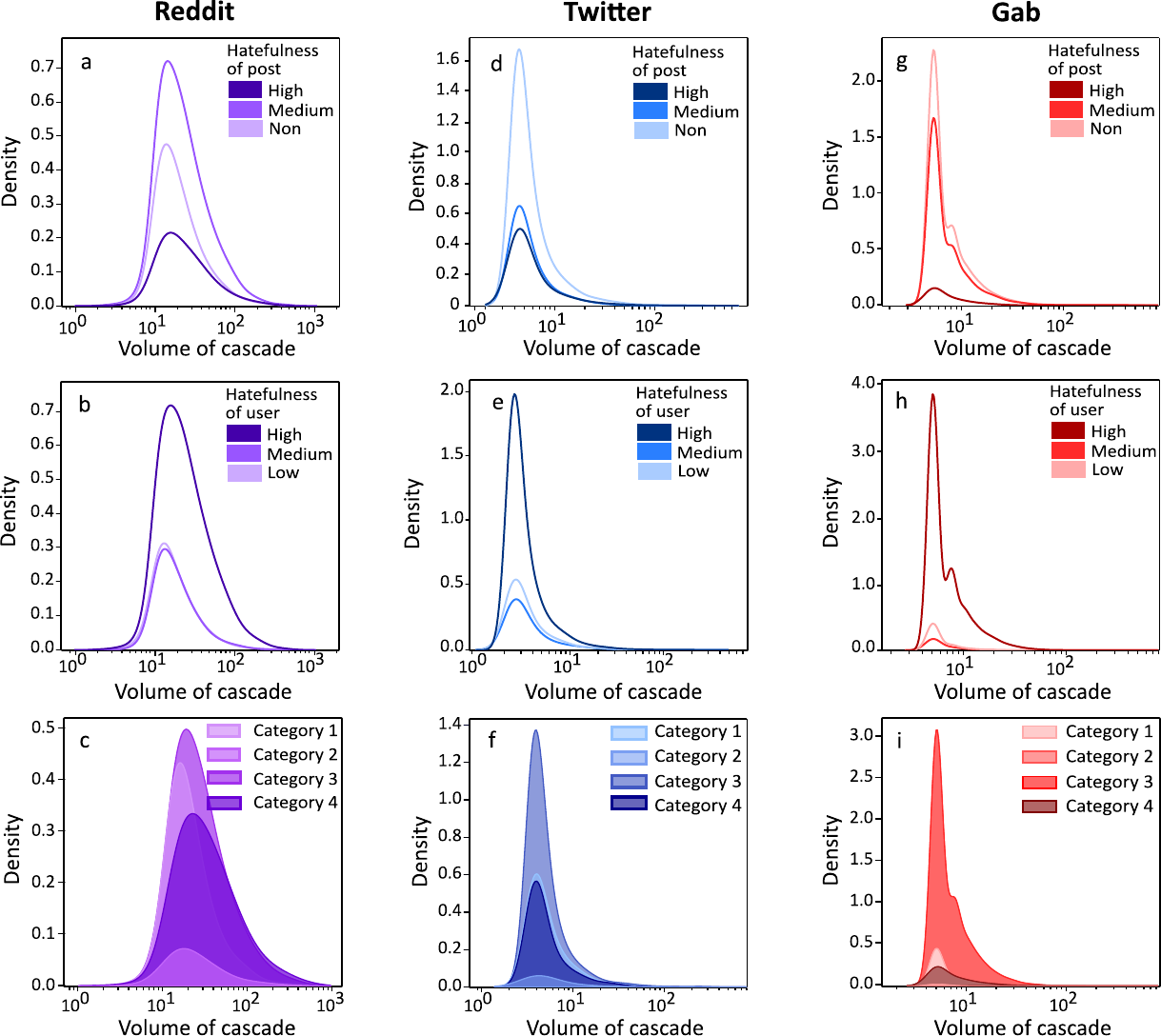}
  \caption{{\bf Hateful users are more potent information spreaders.} Volumetric density distribution of cascades originating from posts of different hatefulness (subplots {\bf a.}, {\bf d.}, {\bf g.}) vs. posts from users with different hatefulness (subplots {\bf b.}, {\bf e.}, {\bf h.}) are presented; for a given value of cascade volume on the $x$-axis, the corresponding $y$-value denotes the density of cascades corresponding to that volume; alternatively, the area under the curve for certain $x$-interval denotes the fraction of cascades having volume within that interval. For all three networks, posts from highly-hateful users are more likely to produce larger cascades. We further analyze the volumetric distribution of cascades originating from hateful users segregated on the basis of hateful posts (subplots \textbf{c.}, \textbf{f.}, \textbf{i.}). Here, \textbf{Category 1} represents a non-hate post from a low-hate user, \textbf{Category 2} represents a high-hate post from a low-hate user, \textbf{Category 3} represents a non-hate post from a high-hate user, and \textbf{Category 4} represents a high-hate post from a high-hate user. In all three networks, low-hate content posted by highly-hateful users tend to breed largest cascades.}
  \label{fig:cascade_combined}
\end{figure*}

\subsection*{Cascade Characterization} We define the cascade formation based on the predominant mode of user interactions over these platforms. For Reddit, a cascade is an $n$-ary tree, originated by a submission and formed by the recursive commenting against the source submission or other comments. For Gab and Twitter, we consider the retweet/quotation activity to be the progenitor of the cascade trees; the original post/submission is treated as the root, while the subsequent reshares are the descendants. We further delineate three quantifiable, structural properties of a cascade -- (i) {\em volume} of a cascade is the sum of out-degrees for all the nodes in the cascade, (ii) {\em width} of a cascade is the maximum number of cascade participants all at the same distance from the root, and (iii) {\em height} of a cascade is the maximum distance from the root to a leaf in the cascade tree. 
% {\color{red} While the volume is an overall indicator of the exposure a specific source post has received, the latter two properties unravel further subtleties of that exposure. A wider cascade is formed with many users interacting with the source content {\em simultaneously}, often indicating a general surge in popularity. On the other hand, a cascade with a greater height signifies a continuous discourse among the cascade participants -- users commenting back and forth to form a discussion or a domino-like sharing/quoting among users connected to each other. }

\subsection*{Detection of Echo Chambers}
\label{sec:detection_ec_main}
In a nutshell, a set of users within a social network is said to form an echo chamber if they exhibit homophily and selective exposure to opinions/ideologies \cite{doi:10.1073/pnas.2023301118,Choi2020}. We build upon these two properties to discover the echo chambers. Most existing methods employ indirect signals, such as referring to some URLs that have been already identified for ideological leaning to define the opinion affinity of users and use that to discover echo chambers \cite{Garimella,Choi2020,shin-echo,doi:10.1073/pnas.2023301118}. Since a substantial number of posts/comments either do not contain URLs~\cite{follower} or are not annotated for opinions, such methods are limited in exploring dynamically evolving social interactions. Instead, we directly use the textual content in the posts. We make topical clusters of user content using our customized topic modeling technique (which is inspired by \cite{DBLP:journals/corr/abs-2008-09470})  and extend these clusters to groups of users based on the social interaction network created from the dataset (see {\color{blue} Materials and Methods}, and {\color{blue} {\em SI Appendix} Section \ref{sec:topic_detection_si}} for more details). Our method is completely automated and needs no external annotation of any sort. 
%The other methodological details are presented in \textcolor{blue}{Materials and Methods}.

\section*{Results}

Our analyses of user influence to spread hate revolve around three primary characterizations of user interactions -- How do different users spread information via cascading along the network? How do hateful users organize themselves in the network? How is the formation of echo chambers entwined with hateful behaviors? 
\begin{figure*}[!t]
    \centering
    \includegraphics[width=\textwidth]{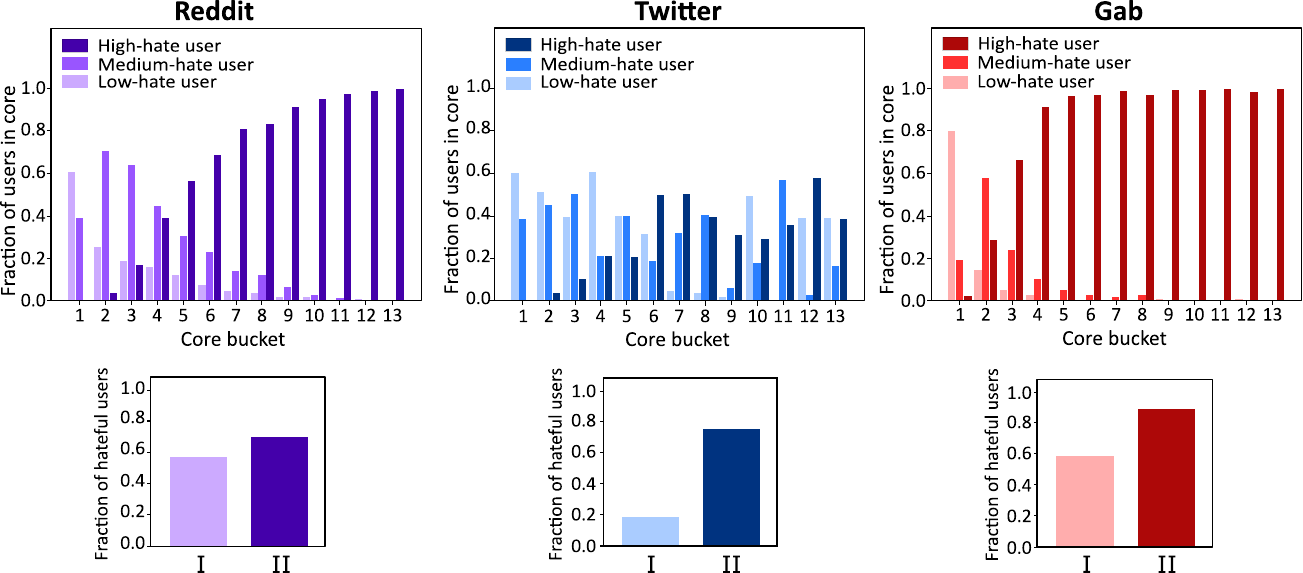}
    \caption{{\bf Hateful users seek a greatly connected position in the network with cohesion.} ({\bf Top}) Distribution of  hatefulness of users at different order cores (a higher core number indicates greater connectivity of a node). The $i${th} value on the $x$-axis represents the $i${th} bucket of $3$ cores each (for example, the core bucket $6$ represents the user nodes having core values $10$-$12$). For all three networks, users with higher core values are increasingly hateful, signifying a greater connectedness to other nodes in the social networks achieved by the highly-hateful users. ({\bf Bottom}) Fraction of (high- and medium-) hateful  users participating in the cascade originated by (I) low-hateful users vs. (II) high-hateful users. A significant affinity is observed among hateful users, pointing towards a possible cohesion among them.}
    \label{fig:core-decomp}
\end{figure*}

\begin{figure}
\centering
\includegraphics[width=\textwidth]{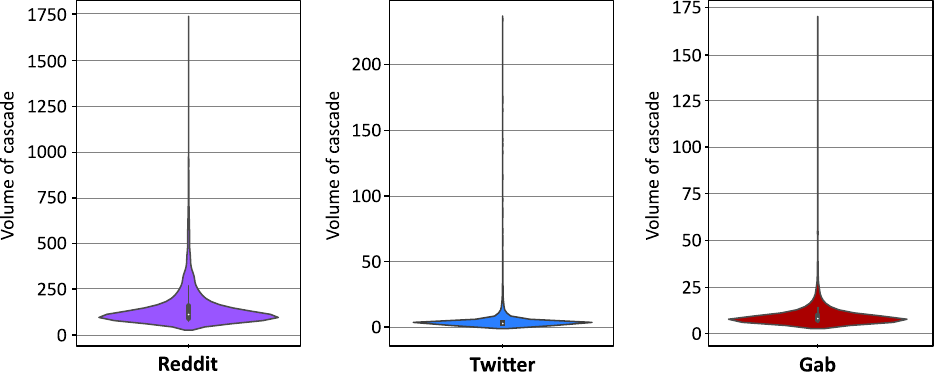}
\caption{Distribution of the volume for the top 10\% volumetrically largest cascades across all three social networks.}
\label{SI:size_dist_violin}
\end{figure}

We start with comparing the volumes of the cascades for different degrees of hatefulness (high, medium, and low) of the source posts and the users who posted them (see {\color{blue} {\em SI Appendix}, Figure \ref{SI:size_dist_violin}} for the overall distribution of volume of cascades for the three datasets). Since the cascade volume comes in a distributed array of values, we show the density distribution over different values of volume.  Figures~\ref{fig:cascade_combined}(b), ~\ref{fig:cascade_combined}(e), and ~\ref{fig:cascade_combined}(h) suggest that when a highly-hateful user writes a post over any of these three networks, the resulting cascade is likely to reach a higher number of users, the Kolomogorov-Smirnov (KS) statistics being $0.198$ for Reddit ($p$-value $<0.001$), $0.147$ for Gab ($p$-value $<0.001$), and $0.049$ for Twitter ($p$-value $=0.02$) (see \textcolor{blue}{{\em SI Appendix}, Section} \ref{sec:significance-testing} for more details related to statistical significance tests). 
% For example, in the case of Reddit, \textcolor{black}{91\% of cascades from low-hate users are contained within a threshold cascade size of 500, while it is just 67\% for highly-hateful users.} This trend is even more dominant in case of Twitter and Gab. While $100\%$ of the cascades formed within Gab from low- or mid-hateful users can be found within a threshold cascade size of \textcolor{black}{70}, this threshold covers only \textcolor{black}{57\%} of the cascades that highly-hateful users started.
To paint a picture of the trend, in Reddit, over $70\%$ of the cascades with volumes more than $50$ originate from high-hate source users; this number rises to $78\%$ if we set the lower bound on cascade size to be $350$. In the case of Gab, such dominance of high-hate users is even more prominent; they are responsible for $88\%$ of the cascades with size $>50$. Twitter, albeit less prominently compared to Gab or Reddit, exhibits a similar trend with $56\%$ of the cascades having volume $>50$ coming out of posts from high-hate users.
This correlation between hatefulness and cascade size is not directly observed if we move from the hatefulness of the source users to that of source posts (see Figures~\ref{fig:cascade_combined}(a), \ref{fig:cascade_combined}(d), and \ref{fig:cascade_combined}(g)). {\color{black}We did not find any significant difference in the cascade volume distributions when categorized by the hate-intensity of the source post in the case of Twitter or Gab (KS statistics are found to be insignificant with $p$-value $>0.05$); for Reddit, we find a weak yet significant difference (KS statistic being $0.11$ with $p$-value $<0.001$).} 

We observe further nuances when the degree of information dissemination is compared for different types of source content posted by hateful and non-hateful users (Figures~\ref{fig:cascade_combined}(c), \ref{fig:cascade_combined}(f), and \ref{fig:cascade_combined}(i)). In case of Twitter and Gab, the density distribution characteristics exhibited by hateful users (Figure~\ref{fig:cascade_combined}(e) and \ref{fig:cascade_combined}(h)) are largely composed of cascades originated by non-hateful posts ({\bf Category 3} in Figures~\ref{fig:cascade_combined}(f) and \ref{fig:cascade_combined}(i)). On both these platforms, non-hateful posts from hateful users are more likely to form larger cascades than even hateful posts from the same category of users. Even on Reddit, if the source user is hateful, the volumetric growth of the cascades does not vary much with the degree of hatefulness of the posts that originated them ({\bf Category 3} and {\bf Category 4} in Figure~\ref{fig:cascade_combined}(e)). Upon additional analysis, we find that this influence of hateful users is not limited to the volume of the cascade but the fraction of hateful interactions as well. In case of Reddit and Twitter, the fraction of hateful interactions is more when the source user is hateful compared to when it is not (see \textcolor{blue}{{\em SI Appendix}, Figure} \ref{SI:fraction_hateful_reply}; KS statistic being $0.869$ for Reddit, $0.878$ for Twitter, and $0.948$ for Gab, all with $p$-values $<0.001$). For Gab, while the absolute number of hateful interactions is higher for hateful users, the fraction gets skewed due to the larger cascades in this case. 

\textcolor{black}{A more nuanced introspection of the interplay between the hatefulness of the users, posts, and cascade growths can be done upon topic-wise analysis. In {\em SI Appendix} (Figure \ref{fig:topic_analysis_response}), we show the distribution of hatefulness and cascade volume for different most-occurring topics across the three platforms. It can be readily seen that the point of hatefulness concentration (low, medium, or high) varies across different topics for different platforms, pointing towards the topical dependence of hateful behavior observed in prior studies~\cite{9458789}. For topics related to anti-abortion or pro-life, we see that the distribution obtains a peak in Gab but the same is not observed for Reddit. This observation points towards the political inclination of users posting on the respective platforms, and the content they like to engage with.}

\textcolor{black}{An apposite question that one may ask at this point is whether or not the role of user hatefulness in information dissemination is actually independent of other user attributes (e.g., age of the user account~\cite{age-and-follower}, follower count~\cite{follower}, etc.). Since Reddit does not provide such profile-centric metadata, Gab and Twitter are considered for this sanity check (see {\em SI. Appendix}, Section 4.E for more details on user metadata analysis). Hatefulness of the user shares a very low normalized mutual information (NMI) with the follower count (i.e., the out-degree of the user node in the social network of the platform): $0.034$ for Gab and $0.044$ for Twitter. Similar patterns are observed in case of user account age as well: $0.045$ and $0.055$ NMI with user hatefulness in case of Gab and Twitter, respectively. These results signify that the hatefulness of the user is indeed an independent variable and not some artifact manifested by other prominent user attributes in a social network.}

\begin{figure}
\centering
\includegraphics[width=\textwidth]{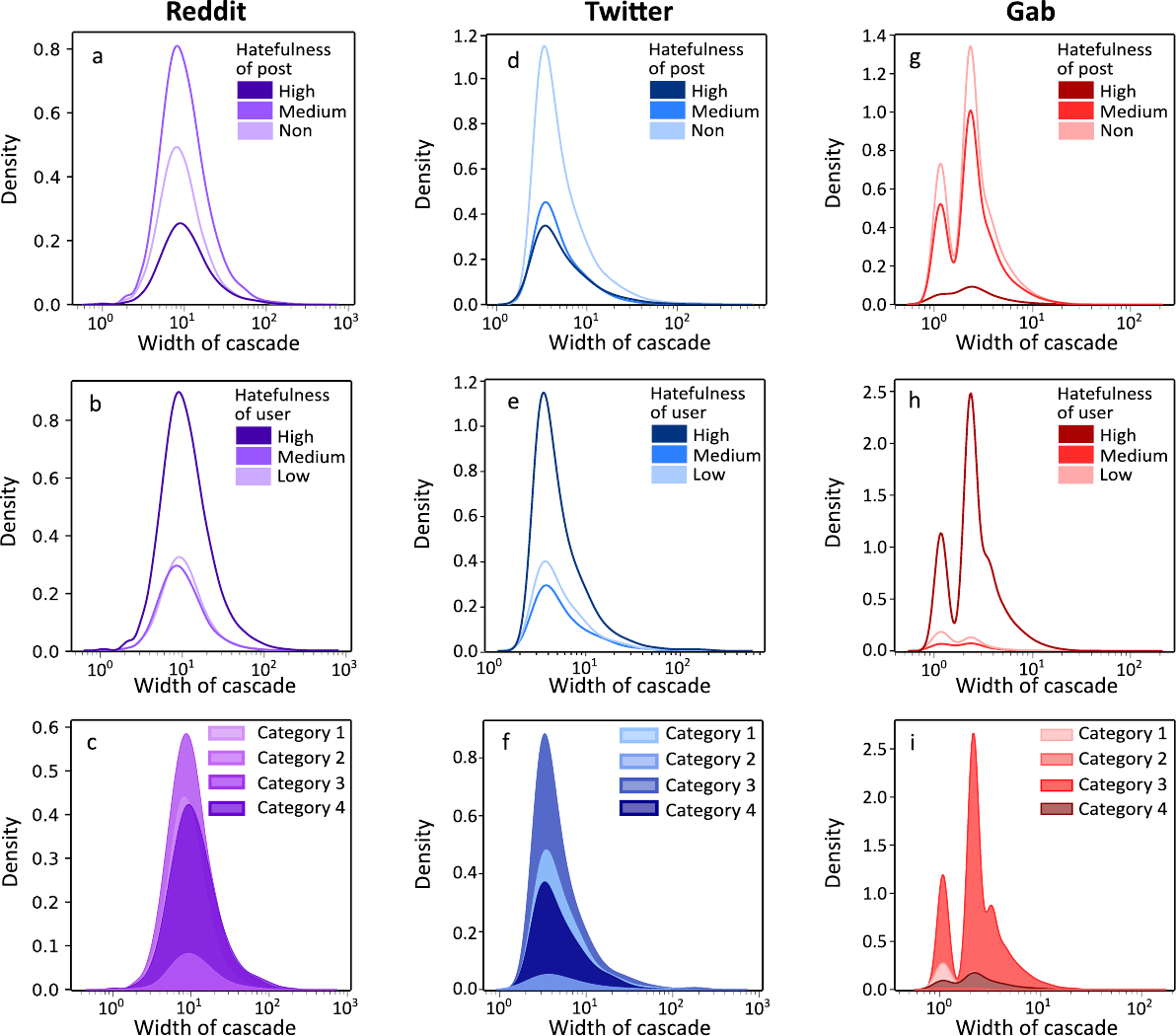}
\caption{Width distribution of cascades originating from hateful posts (subplots {\bf a.}, {\bf d.}, {\bf g.}) vs. posts from hateful users (subplots {\bf b.}, {\bf e.}, {\bf h.}); for a given value of cascade
width in the $x$-axis, the corresponding $y$-value denotes the density of cascades corresponding to that width.
For all three networks, posts from highly-hateful users are more likely to produce cascades of larger width. We further present the width distribution of cascades originating from hateful users segregated on the basis of hateful posts (subplots \textbf{c.}, \textbf{f.}, \textbf{i.}).
Here, \textbf{Category 1} represents a low-hate post from a low-hate user, \textbf{Category 2} represents a high-hate post from a low hate user, \textbf{Category 3} represents a low-hate post from a high-hate user, and \textbf{Category 4} represents a high-hate post from a high-hate user. In all three networks, low-hate content posted by highly-hateful users tend to breed cascades with largest width.}
\label{fig:cascade_width_distr}
\end{figure}

\begin{figure}
\centering
\includegraphics[width=0.67\textwidth]{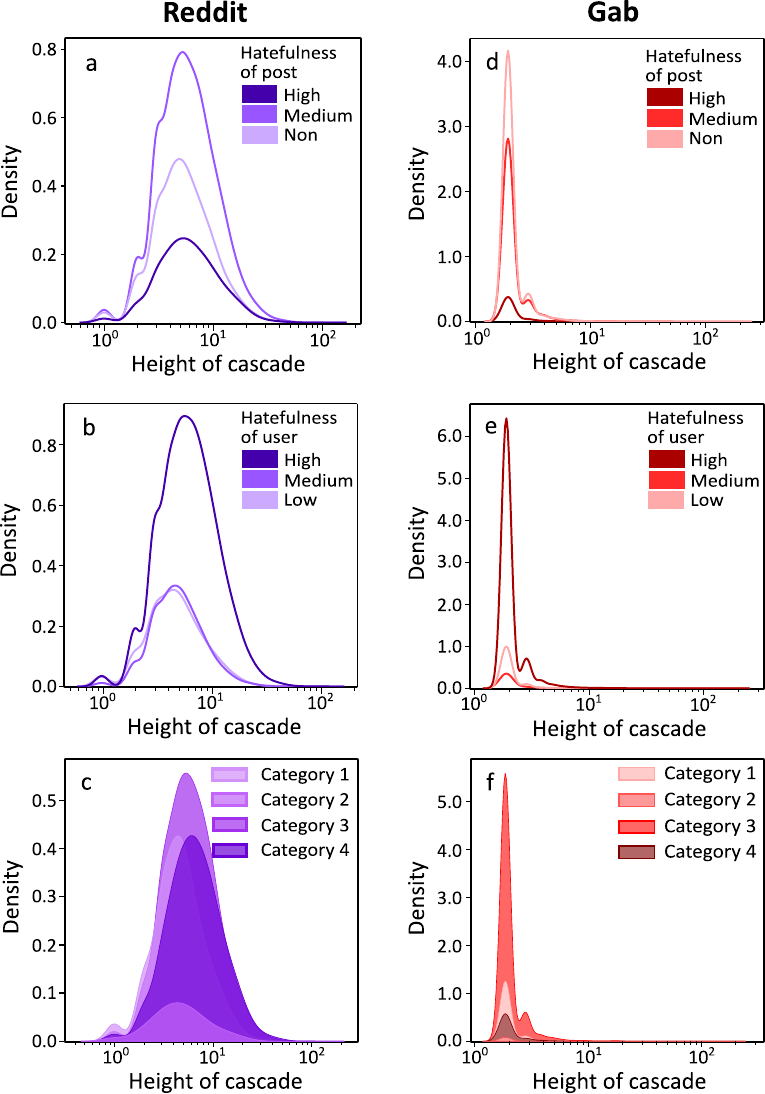}
\caption{
Height distribution of cascades originating from hateful posts (subplots \textbf{a.}, \textbf{d.}) vs. posts from hateful users (subplots \textbf{b.}, \textbf{e.}); for a given value of cascade
height in the $x$-axis, the corresponding $y$-value denotes the density of cascades corresponding to that height.
For both Reddit and Gab, posts from highly-hateful users are more likely to produce cascades of larger height. We further present the height distribution of cascades originating from hateful users segregated on the basis of hateful posts (subplots \textbf{c.}, \textbf{f.}). 
Here, \textbf{Category 1} represents a low-hate post from a low-hate user, \textbf{Category 2} represents a high-hate post from a low hate user, \textbf{Category 3} represents a low-hate post from a high-hate user, and \textbf{Category 4} represents a high-hate post from a high-hate user. In all three networks, low-hate content posted by highly-hateful users tend to breed cascades with largest height.
We do not present analysis of the height distribution of cascades for Twitter, since the dataset that we use does not show cascade height of more than $3$.}
\label{fig:cascade_height_distr}
\end{figure}

We further analyze the structural properties of the cascades in terms of the width and height of the cascade tree \textcolor{blue}{(see {\color{blue}{\em SI Appendix}, Figures \ref{fig:cascade_width_distr}, \ref{fig:cascade_height_distr})}}. 
% Generally, the height of a cascade tree (length of the path from source node to the most distant leaf node) is reflective of continuous discourse between the cascade participants, whereas the width (maximum number of nodes sharing the same depth) is more indicative of a rapid growth of information dissemination (since at any time, the width of a cascade tree indicates the number of users participated in the cascade independently). 
We observe a positive dependence, although in degrees that are platform-dependent, between hatefulness of the source user and the height of the cascade (see {\color{blue}{\em SI Appendix}, Figure \ref{fig:cascade_height_distr}}) as well as with the width of the cascade (see {\color{blue}{\em SI Appendix}, Figure \ref{fig:cascade_width_distr}}).
Altogether, even though {\em hate attracts hate} remains true for all three networks, {\em who is posting} plays an even more dominating role.
This is a crucial observation since most prescriptions dealing with hate speech strongly emphasize regulating hateful content; however, a hateful user, if not unchecked, is more likely to disturb the overall harmony compared to isolated instances of hateful content.

If the actors play a more pivotal role in diffusing information over a network than the content itself, our natural intuition will point towards enquiring about the network organization; highly-hateful users should organize themselves in a way that maximizes their influence. This hypothesis is readily justified when we investigate the distribution of hateful users among different depths of the $k$-core decomposition, as shown in Figure~\ref{fig:core-decomp} (top) (see {\color{blue}{\em SI Appendix}, Section \ref{sec:k-core_decomposition}} for the details about the $k$-core decomposition). Across all three networks, nodes with higher $k$-core numbers are more likely to be high-hateful (Spearman $\rho$ values being $0.68$ for Reddit, $0.30$ for Twitter, and $0.77$ for Gab, all with $p$-value $<0.001$). Reddit and Gab exhibit a monotonic increase in highly-hateful users as the core number increases. For Reddit, highly-hateful users tend to dominate the distribution once we surpass $15$-cores. For Gab, this transition comes at an even shallower level, probably due to the domination of hatemongers that Gab is infamous for. Twitter, albeit showing a similar overall trend, has some fluctuations in the hatefulness distribution; high- and mid-hateful users start dominating after $18$-cores in this case. This organization is reflected in the interaction pattern of the hateful users as well -- strong connectivity among the hateful users indicates that they would disseminate information together as well. The cascades initiated by posts from hateful users are more likely to attract other hateful users to participate compared to cascades initiated by non-hateful users (Figure~\ref{fig:core-decomp} ({bottom})), though in this case, Twitter shows the most disparity in user engagement \cite{rathje2021out}, with Reddit being the closest to a balanced scenario among the three. These results also confirm previous studies pointing out the existence of collaborative networks among hate-spreaders \cite{Johnson2019-hate-network, hate-spreader-theory}.

\begin{figure*}[!t]
    \centering
    \includegraphics[width=\textwidth]{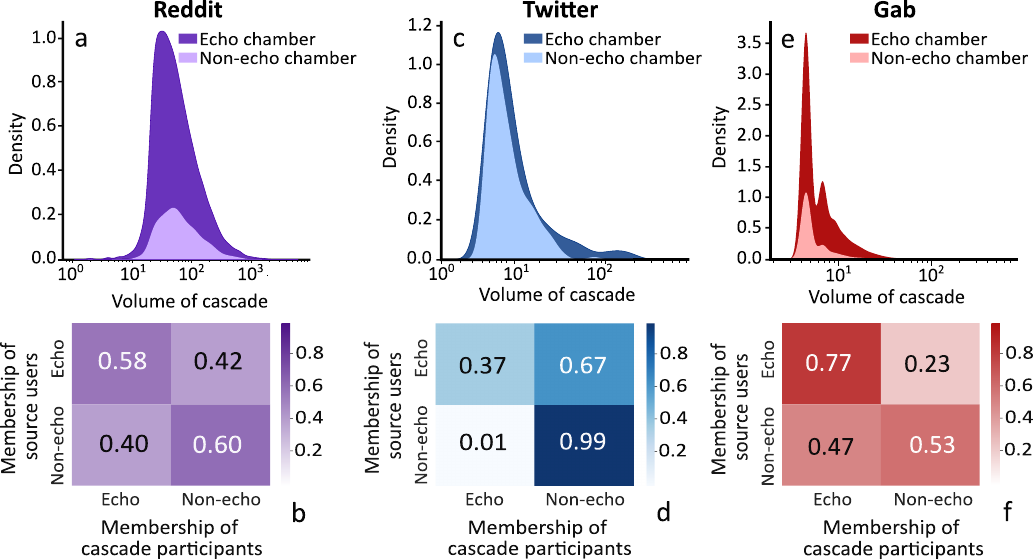}
    \caption{{\bf Echo chambers amplify information spreading potential of hatemongers.} For each network, {\bf a.}, {\bf c.}, and {\bf e.} separate out the density distribution of cascade volume for highly-hateful source users when they do or do not belong to any echo chamber; particularly for Reddit and Gab, echo-chambered users are responisble for the majority of the high-volume cascades from highly-hateful users. {\bf b.}, {\bf d.}, and {\bf f.} compare the fraction of interactions in the cascades originated by highly-hateful echo chamber members coming from other members vs. non-members; here, in Gab and Twitter, echo chambered users are more likely to participate in cascades originated by fellow echo chamber members.}
    \label{fig:echo-cascade}
\end{figure*}

The previous observation from core decomposition and user engagement characterization points to the fact that hateful users exhibit a unique cohesion. What is left is to verify whether this cohesion can be related to polarization or not. To this end, we employ our proposed method of discovering echo chambers to unfold the further nuances of interactions materialized by the users. We start by dissecting our initial results on cascade growth dynamics (shown earlier in Figure~\ref{fig:cascade_combined}), now aware of the presence of the echo chambers in Figure~\ref{fig:echo-cascade}. Whether or not a highly-hateful source user is a member of some echo chambers largely determines the growth disparity occurring among cascades; even highly-hateful users who are not members of any echo chambers procure a cascade growth very similar to those originated by low- or medium-hateful users. In all three networks, the density distributions of cascade volumes corresponding to highly-hateful users are very close to the fraction of those which have originated by highly-hateful users {\em within} echo chambers. The $p$-value for KS statistic between the cascade volume distributions from all highly-hateful source users and those who belong to echo chambers comes to be $>0.05$ for all three networks, thereby accepting the null hypothesis that these two distributions are indeed the same. We observe significant disparity among the cascade constituent users as well. In the case of Reddit, for example, if the source user is a member of an echo chamber, then $58\%$ of the cascade participants are coming from echo chambers as well; if the source user is not a member of an echo chamber, this number drops to $40\%$ (see Figure~\ref{fig:echo-cascade}(b). This disparity is even sharper in the case of Twitter and Gab, as shown in Figures~\ref{fig:echo-cascade}(d) and \ref{fig:echo-cascade}(f) This is pretty much at par with what we found about the cohesion of hateful users in Figure~\ref{fig:core-decomp} ({bottom}) --  hate-spreaders are seen to exhibit more affinity to participate in cascades originated by fellow hate-spreaders on both Twitter and Gab compared to that on Reddit.  
%Upon probing the structural properties of these cascades again with the height and width distributions, we can further observe the role of echo chambers in cascade growth. Cascades originated by members of echo chambers tend to grow more in height compared to those with source users not a member of any echo chamber (see {\color{blue}{\em SI Appendix}, Figure \ref{fig:cascade_height_echo}}); on the other hand, there is not much observable difference between the two cases in terms of the width of the cascade (see {\color{blue}{\em SI Appendix}, Figure \ref{fig:cascade_width_echo}}). The difference in cascade volume distribution resulting from the echo chamber membership can therefore be linked to the fact that users within echo chambers are more likely to engage in continuous discourse and push the cascade to grow more in height; the width distribution remains almost identical, possibly signaling that the number of parallel engagements over the network is not much related to echo chamber formation.

\begin{figure*}[!t]
    \centering
    \includegraphics[width=\textwidth]{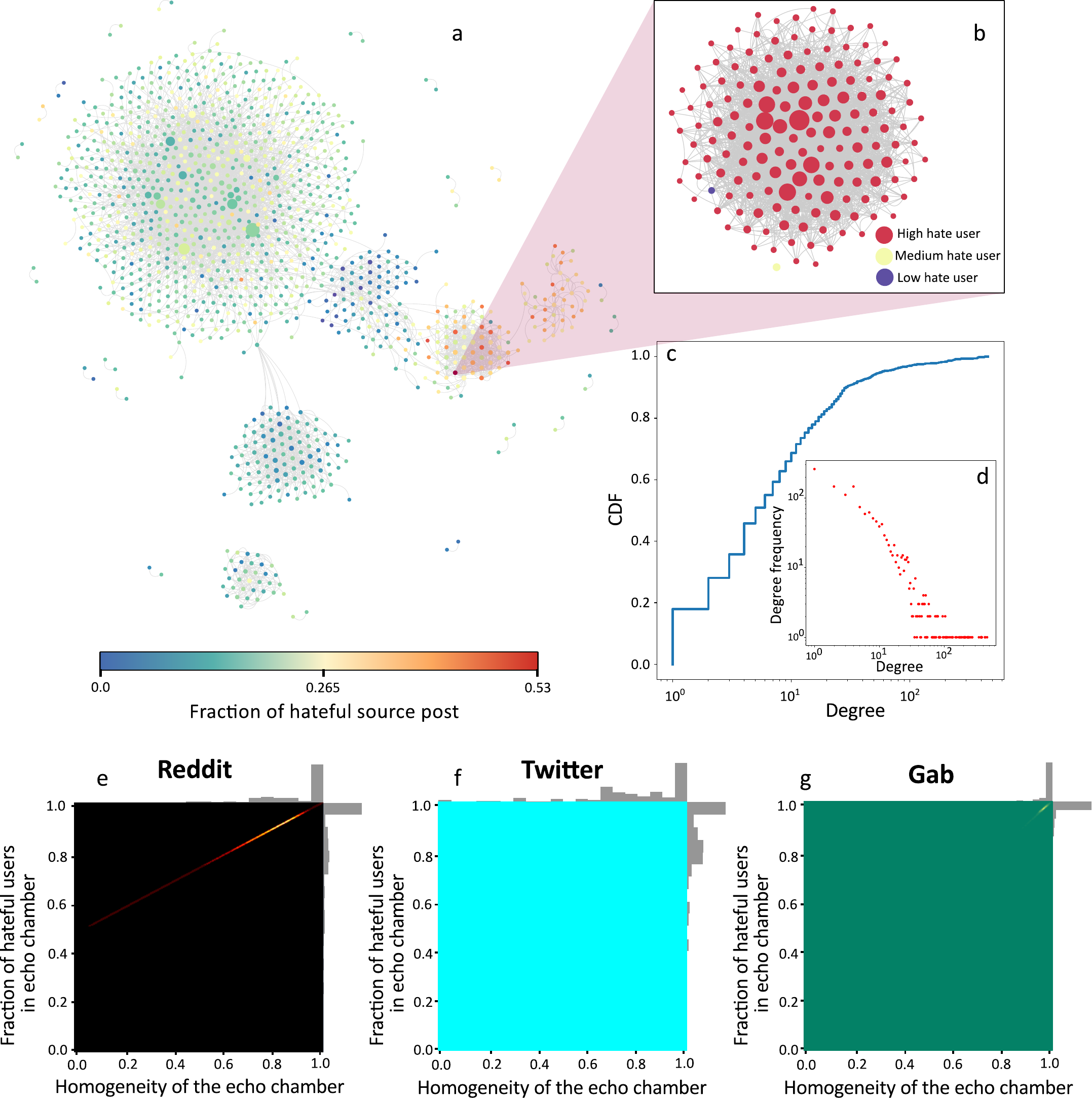}
    \caption{{\bf Hatemongers dominate echo chambers.} {\bf a.} A sample distribution of echo chambers in Reddit; each node represents an echo chamber color coded with the fraction of hateful source posts; an edge between two echo chambers denotes shared users (see similar example networks for Gab and Twitter in {\color{blue}{\em SI Appendix}, Figures \ref{fig:gab_network} and \ref{fig:twitter_network}}). {\bf b.} A user interaction network within an example echo chamber; each node being a user with edges defined by {\em reply-to} interactions. {\bf c.} and {\bf d.} show the degree distributions of the network shown in {\bf a.} The distributions of homogeneity vs. fraction of hateful users in echo chambers for Reddit, Twitter, and Gab, respectively are shown in {\bf e.}, {\bf f.}, and {\bf g.} One would expect a triangular distribution of the echo chambers in the heat maps if there were no relation between the homogeneity and fraction of hateful users present in the echo chambers. Instead, homogenous echo chambers are dominated by hateful users in all three networks.}
    \label{fig:echo-chamber-purity}
\end{figure*}

Up to this point, our findings divulge key roles of hateful users in a social network and how further complexities are integrated into this context with the formation of echo chambers. A sample network of echo chambers (nodes are echo chambers, and edges signify shared users) in Reddit is shown in Figure~\ref{fig:echo-chamber-purity}(a); the degree distribution characteristics of the same network are presented in Figures~\ref{fig:echo-chamber-purity}(c) and \ref{fig:echo-chamber-purity}(d). The inner organization of the users within an example echo chamber is shown in Figure~\ref{fig:echo-chamber-purity}(b). Finally, we seek to characterize the echo chambers based on their constituent users. In the context of hate speech, we first put forward a simple definition of  {\em homogeneity} of an echo chamber as:
\begin{equation*}
    p(E) = \frac{\lvert C( E_\text{H}) - C( E_\text{N} ) \rvert}{C( E )}
\end{equation*}
where $E$ is an echo chamber described as a set of users; $E_\text{H}$ and $E_\text{N}$ are the set of high/mid-hateful and low-hateful users in $E$, respectively; $\lvert \cdot \rvert$ denotes absolute value of scalar; and $C( \cdot )$ denotes the size of a set. In Figures \ref{fig:echo-chamber-purity}(e), \ref{fig:echo-chamber-purity}(f), and \ref{fig:echo-chamber-purity}(g), we plot the variation of the hatefulness of echo chambers with their homogeneity. Generally, we can observe a linear relationship between the two for all three networks. However, the homogeneity distribution of echo chambers is different for different platforms. In Reddit and Twitter, we can find echo chambers residing towards the least homogenous  end of the spectrum; the echo chambers inGab are predominantly skewed towards the most homogenous (and, as a result, most hateful) end. This again is expected given the notoriety of Gab for giving a free pass to hate spreaders \cite{hate-gab-1}.

\section*{Conclusion}

We presented a comprehensive analysis of hateful behavior on online social networks under the beacon of information dissemination and polarization in terms of echo chamber formation. Upon investigating three popular social networking sites, namely, Reddit, Twitter, and Gab, we observed multiple intriguing patterns of the spread of hate speech. We established that once posted on a network, hateful content tends to provide an assembly point for further hateful behavior; however, the hatefulness of a user plays a more dictating role in this regard compared to a single content that is hateful. Within all three networks, posts from hateful users tend to engage more users over time, both in terms of total volume as well as the degree of hatefulness, even when the source post is apparently non-hateful. This corroborates previous claims that the combat mechanisms of online hate should not be centered around content moderation only.

Our findings suggest that the observed influence of hateful users in terms of cascade growth can be linked to the fact that they are usually placed within the deeper cores of the networks indicating a greater connectedness with other users. We observed cohesion among the hateful users, the degree of which varies for different networks. To check whether this cohesion results from the organized activity, we proposed a fully-automated method to detect echo chambers in networks based on selective exposure to ideology and homophily. Not to our surprise, we found that the cascade growth dynamics linked to highly-hateful users consist of source users belonging to some echo chambers. Following the echo chamber normative, these cascades, in turn, largely consist of participant users from echo chambers. Across all three platforms, we found that echo chambers are strongly biased towards hateful members -- the homogeneity of echo chambers, defined as the degree of mixing between hateful and non-hateful users, is highly skewed towards the hateful end. To the best of our knowledge, this is the very first empirical validation of previously-made conjectures linking hate-mongering to echo chambers \cite{cinelli2021dynamics,ec_moral_radicalism}. These findings might point toward a shortcoming realized by the content recommendation algorithms of most current social networking platforms. Most of the popular online social networking platforms promote content based on their popularity, reinforcing their already acquired popularity. Hatemongers might be able to exploit this by their strategy to popularize content. Since the boost in the influence of such users is primarily driven by echo chambers, a possible countermeasure can be devised based on conditional content promotion by taking the presence of echo chambers into account.

Our findings primarily connect the existing studies on the recently increasing trend of polarized behavior online and the diffusion of hateful content on the Internet. While we correlated the internal organization of hatemongers and observed disparities in information dissemination, this does not necessarily establish a causal relationship. A chronological observation of echo chamber formation and the evolution of hatefulness over the network may provide the community with the answer to the question: Does hate drive the formation of echo chambers or vice versa? Similar studies may provide further insights into other menaces of online social networks, such as toxicity and misinformation.

\section*{Materials and Methods}
Here we provide further details on the dataset acquisition, the availability of the data, information about the composition, and other methodological details.
\subsection*{Data Acquisition}
For Reddit, we used the \href{https://github.com/pushshift/api}{Pushshift APIs} to fetch data across the year 2019 and filtered out selected subreddits. The description of the subreddits are provided in \textcolor{blue}{{\em SI Appendix}, Table \ref{tab:subreddit_detail}}.
For Gab, we wrote our own custom scraper to fetch data \textcolor{black}{from October, 2020 to September, 2021}. The systematic scrapping method is described in \textcolor{blue}{{\em SI Appendix}, Section \ref{sec:dataset_si}.\ref{sec:gab_dataset}}.
For Twitter, we used the same mechanism to extract data as mentioned in \cite{Garimella}. We used the Twitter dump available at \href{https://archive.org/details/twitterstream}{The Twitter Stream Grab}. \textcolor{black}{We collected data within the period of April to June, 2019} and then filtered data based on hashtags (more details in \textcolor{blue}{{\em SI Appendix}, Table \ref{tab:twit_hashtag_detail}}).

\subsection*{Data Availability}
The data for Reddit can be extracted through the publicly-accessible \href{https://github.com/pushshift/api}{Pushshift APIs}. We also make the anonymized version of the dataset for all three networks available on the Open Science Framework \href{https://doi.org/10.17605/OSF.IO/6AZCF}{(10.17605/OSF.IO/6AZCF)}. %\cite{Goel_Sahnan_Dutta_Bandhakavi_Chakraborty_2022}. 
Further description of the dataset is provided in the following sections. The source codes for the analyses conducted in the paper are uploaded on the following Github repository:  \href{https://github.com/LCS2-IIITD/Hate-mongerer-and-echo-chambers}{https://github.com/LCS2-IIITD/Hate-mongerer-and-echo-chambers}.

% The data for Reddit was extracted through the publicly-accessible \url{https://github.com/pushshift/api} APIs. As for Gab and Twitter, we make the data available according to their Terms of Services. Anonymized version of the dataset has been deposited in Open Science Framework  \href{https://doi.org/10.17605/OSF.IO/6AZCF}{(10.17605/OSF.IO/6AZCF)} \cite{Goel_Sahnan_Dutta_Bandhakavi_Chakraborty_2022}. Further description of the dataset is provided in the following sections.

\subsection*{Empirical Datasets}

\if 0
\begin{table}[h]
    \centering
    \begin{tabular}{ c|c|c|c|c } 
     \hline
     \textbf{Platform} & \textbf{\# posts} & \textbf{\# users} & \textbf{\# cascades} &\textbf{Duration}\\
     \hline
     \textbf{Reddit} & 851676 & 97349 & 22146 & Jan'19-Dec'19 \\
     \textbf{Twitter} & 31500556 & 6709951 & 59638 & Apr'19-Jun'19 \\
     \textbf{Gab} & 322561 & 29066 & 22978 & Oct'20-Sept'21 \\
     \hline
    \end{tabular}
    \caption{Statistics of datasets. We report the number of posts, number of users, number of cascades, start date for collection of data, and duration of the data collection.}
    \label{tab:dataset_table}
    \vspace{-5mm}
\end{table}
\fi

We report the statistics and summary of the dataset in \textcolor{blue}{{\em SI Appendix}, Table} \ref{tab:detail_dataset}. 
The three networks do not index posts into topics by default. For Reddit, we filtered the topics of interest initially using subreddits (see {\color{blue}{\em SI Appendix}, Table \ref{tab:subreddit_detail}}). For Twitter and Gab, we did the same through the use of hashtags and keywords (see {\color{blue}{\em SI Appendix}, Tables \ref{tab:twit_hashtag_detail} and \ref{tab:gab_keyword_detail}}, respectively). The focus of topic selection was to ensure opinionated content and interactions; a good proportion of these topics had been used by previous works as well (e.g., gun ownership, abortion, etc.) \cite{Garimella, doi:10.1073/pnas.2023301118}.
The topics also provide a good interest as some of them also have alignment with a user's political opinions as shown previously by \cite{doi:10.1073/pnas.2023301118}. For Twitter and Gab, we created links between posts by fetching posts and their retweets, whereas, for Reddit, we constructed the network by fetching the submissions and their corresponding comments. \textcolor{blue}{{\em SI Appendix}, Section \ref{sec:dataset_si}} contains a detailed description of each dataset along with the hate scoring, characterization methods. \textcolor{black}{The section also includes our approach to validate our hate scoring method, for which we employed three annotators to annotate the hatefulness of the content, and then compared their annotations with the output of our method.}

\subsection*{Detection of Echo Chambers}
We propose an automated method for the detection of echo chambers in a social network, wherein we exploit the recent advancements in Natural Language Processing to our advantage - getting rid of any kind of annotations earlier methods required \cite{doi:10.1073/pnas.2023301118, Choi2020}. We use a state-of-the-art sentence encoder \cite{cer2018universal} over each piece of content and utilize its capability to generate semantically-accurate embeddings to cluster similar content. Since the dimension of these real-valued embedding vectors is huge, we apply dimensionality reduction methods to obtain embeddings of smaller sizes. We then perform clustering over these reduced embeddings to extract groups of posts that are essentially discussing a similar theme. The groups or clusters of posts can also be understood as topics. As mentioned earlier, %{\color{blue}{\em Characterizing Hate and Echo Chambers: Detection of Echo Chambers}}, 
echo chambers exhibit two properties, \textit{ideological homophily} and \textit{selective exposure}; we enforce ideological homophily through the inherent nature of the advance sentence encoder to capture the semantics of a sentence; furthermore, we consider two users to be a part of an echo chamber if and only if they have posted/interacted in multiple common topics on their own accord, which satisfies selective exposure. A detailed description of the method for echo chamber discovery can be found in {\color{blue}{\em SI Appendix}, Section \ref{sec:topic_detection_si}}.

\section*{Acknowledgements}
We thank Logically for their precious insights and financial support in developing this study. We also thank H. Russell Bernard, Rula Awad, Sarah Masud, Srishti Sahni, Rahul Kukreja, Megha Sundriyal, and Debarka Sengupta for their valuable suggestions.

\newpage
\onecolumn

%% Adds the main heading for the SI text. Comment out this line if you do not have any supporting information text.
{\Huge{\bf Supplementary Materials}}\vspace{3mm}

Here we provide details of dataset acquisition, hate scoring mechanism, model specifications, $k$-core decomposition and significance testing. We also present some additional results supplementing those shown in the main text.

\section{Dataset}
\label{sec:dataset_si}
In this section, we list out specifics for data collection and filtration, their composition, and our hate scoring/characterization methodology. Table~\ref{tab:detail_dataset} contains a detailed description of the three datasets and their user-interaction networks.

\begin{table}[ht]
    \centering
    \begingroup
    \setlength{\tabcolsep}{7pt} % Default value: 6pt
    \renewcommand{\arraystretch}{1.2}
    \begin{tabular}{ l|l|r|r|r|r|r } 
     \textbf{Platform} & \textbf{Duration} & \textbf{\# posts} & \textbf{$C(U)$} & \textbf{$C(I)$} & \textbf{$\langle k\rangle$} & \textbf{$C(CAS)$}  \\
    %  \textbf{Platform} & \textbf{\# posts} & \textbf{\# users} & \textbf{\# cascades} &\textbf{Duration}\\
     \hline
     Reddit & Jan'19-Dec'19 & 851002 & 97349 & 464087 & 9.53 & 22146\\
     Gab & Oct'20-Sept'21 & 322561 & 29066 & 120211 & 8.52 & 22978\\
     Twitter & Apr'19-Jun'19 & 31500556 & 6709951 & 15105559 & 4.50 & 59638\\\hline
    \end{tabular}
    \endgroup
    \caption{Detailed statistics of the datasets. We list, for each dataset, the \textbf{duration} of its collection and the \textbf{number of posts} in the final dataset; $U$, $I$, and $\langle k \rangle$ denote the set of all users, the set of all unique user-user interactions, and the average degree, respectively, in the constructed user-interaction network, and $CAS$ denotes the set of all post cascades we extract from the dataset. $C(.)$ denotes the size of a set.}
    \label{tab:detail_dataset}
\end{table}

\subsection{Data Collection and Filtration}
Here we explain the data collection process for all three social media networks individually, along with details of their availability. As part of the filtering process, we mapped key political and social events that took place over the duration of the collection of each dataset. For collecting content based on these events, we performed a keyword/hashtag-based extraction for Gab and Twitter, while subreddit titles and descriptions were enough for Reddit. For each platform, we only considered the user-user interactions (submission-comment on \textbf{Reddit}, post-reblog on \textbf{Gab}, tweet-retweet on \textbf{Twitter}) and the textual content of these interactions for our analyses.

\subsubsection{Reddit}
The data dump, extracted from the \href{https://github.com/pushshift/api}{Pushshift API}, contains submissions and their comments from a variety of subreddits and spans the year 2019. We further select subreddits that cater to each side of the socio-political spectrum and contain discussions about major real-world events that took place in 2019. The subreddits selected are explained in Table~\ref{tab:subreddit_detail}.

\begin{table}[ht]
    \centering
    \begingroup
    \setlength{\tabcolsep}{7pt} % Default value: 6pt
    \renewcommand{\arraystretch}{1.2}
    \begin{tabular}{ p{0.2\textwidth}|p{0.75\textwidth} } 
     \textbf{Subreddits} & \textbf{Description}\\
    %  \textbf{Platform} & \textbf{\# posts} & \textbf{\# users} & \textbf{\# cascades} &\textbf{Duration}\\
     \hline
     MensRights, againstmensrights, MensLib & Discussions entailing men's legal rights and societal issues they face in everyday life. \textbf{MensRights} majorly comprises legal rights, but general discussions around their relationship with society are also allowed; \textbf{againstmensrights} is generally based on uncovering hate and toxicity in r/MensRights but also in the general men's rights movements; \textbf{MensLib} is a more general subreddit, created in an effort to allow positive and open-minded discussions on men's issues.\\\hline
     abortion, prolife, prochoice & Discussions comprising issues around abortion and the recent pro-life/pro-choice movement across the world. \textbf{abortion} can be explained as more of a support group for people dealing with abortion and comprises general conversations spanning both spectrums of the pro-life/pro-choice movement; \textbf{prolife} and \textbf{prochoice} subreddits, as their names suggest, contain discussions around the respective sides of the debate.\\\hline
     environment, climatechange,\newline climateskeptics & Discussions around changes in the environment and their corresponding socio-political movements that are taking place across the world. \textbf{environment} contains posts around recent news, information and issues related to the changes in the environment; \textbf{climatechange} comprises rational discussions and the consequences of climate change in the present day and the coming years; \textbf{climateskeptics} is a subreddit majorly focused on uncovering alarmism and conspiracies in recent discussions on environmentalism.\\\hline
     aliens, area51raid, UFOs & Discussions comprised alien life, the famous US Air Force facility and questions about flying object sightings. \textbf{aliens} contains conversations majorly on the possibility of extra-terrestrial life; \textbf{area51raid} comprises posts about conspiracies about the highly classified US Air Force facility in Nevada and events to "storm" it together; \textbf{UFOs} is a subreddit listing public sightings of unidentified flying objects around the world.\\\hline
     conspiracy, TruthLeaks & Discussions around the most famous conspiracies of all time from across the world. \textbf{conspiracy} serves as a thinking ground for any general conspiracy theory and people's opinions on them; \textbf{TruthLeaks} contains open-source investigations and evidence to discuss and uncover some of the major conspiracies in play today.\\\hline
    \end{tabular}
    \endgroup
    \caption{List of subreddits in the Reddit dataset we use, along with brief descriptions for each of them. Subreddits are grouped in the list on the basis of similarity in the topics of their discussions.}
    \label{tab:subreddit_detail}
\end{table}

% \todo{(add some details about number of submissions, comments and the user interaction network)}

\subsubsection{Gab}
\label{sec:gab_dataset}
The Mastodon (an open-source social networking service) based microblogging platform is known for its user's far-right socio-political ideology. For extracting the Gab dataset, we identified a set of popular users with high posting activity aligned with the real-world events that happened between October 2020 and September 2021. These users were then used as seed nodes for a custom scraper that we designed to recursively collect users that follow them. Collecting for multiple hops of follow relations, we then built a large social network of users based on followership. The scraper then extracted posting history from the user timelines. Finally, we filtered the data using keyword-based analysis that aligns with the socio-political events we identified as occurring during the time of our collection. Table~\ref{tab:gab_keyword_detail} explains the keywords that we filtered. The issues identified for filtration majorly comprise topics related to the US politics, with a high percentage of them from the perspective of the far-right supporting population across the world.

\begin{table}[ht]
    \centering
    \begingroup
    \setlength{\tabcolsep}{7pt} % Default value: 6pt
    \renewcommand{\arraystretch}{1.2}
    \begin{tabular}{ p{0.2\textwidth}|p{0.75\textwidth} } 
     \textbf{Keywords} & \textbf{Description}\\
    %  \textbf{Platform} & \textbf{\# posts} & \textbf{\# users} & \textbf{\# cascades} &\textbf{Duration}\\
     \hline
     racism, black, white, arrest, murder & These keywords essentially point to discussions on the prevalence of racism across the United States, fueled by the George Floyd incident; they contain both sides of the debate, i.e., people against racism and white supremacy. \\\hline
     abortion ban, parenthood, texas, prochoice & Discussions majorly over the amendments to abortion laws in various states of the US and their corresponding movements; contain clashes between people from both sides of the debate along with news and information about major incidents. \\\hline
     trump, MAGA, election, biden & Content comprising of the build-up and aftermath of the 2020 US Presidential elections, majorly supporting Donald Trump, indicating the one-sided nature of the platform. \\\hline
     gun laws, ban, shootout & Discussions regarding gun access laws in the US, including the recent movement for supporting the ban of these guns across many states, contain discussions over the various school shootings that took place in the USA. \\\hline
     vaccines, anti-vax, vaxxhappened & Contains opinions of people across the world on the use of vaccines fueled by the COVID-19 pandemic; contains a significant amount of discussions from people identifying themselves as "anti-vaxxers".\\\hline
    \end{tabular}
    \endgroup
    \caption{A brief overview of the keywords in the Gab dataset we use, along with a brief description of each group. Keywords in the list are grouped on the basis of the similarity of the topics they represent.}
    \label{tab:gab_keyword_detail}
\end{table}

\subsubsection{Twitter}
We followed the approach suggested in \cite{doi:10.1073/pnas.2023301118} to extract a data dump from \href{https://archive.org/details/twitterstream}{The Twitter Data Stream}. This data dump is a 10\% snapshot of the international Twitter feed for the months between April 2019 and June 2019, and the content comprises of a variety of topics ranging from social issues to friendly banter. We performed a hashtag-based analysis over this dump and extract only those tweets (and their retweets) that cater to the real-world issues in discussion during the months the dump belongs to. The events extracted include discussions over the US politics, conspiracies, social rights, and others. Table~\ref{tab:twit_hashtag_detail}  details some hashtags/keywords in our dataset and brief descriptions for each of them.

\begin{table}[ht]
    \centering
    \begingroup
    \setlength{\tabcolsep}{7pt} % Default value: 6pt
    \renewcommand{\arraystretch}{1.2}
    \begin{tabular}{ p{0.2\textwidth}|p{0.75\textwidth} } 
     \textbf{Hashtags/Keywords} & \textbf{Description}\\
    %  \textbf{Platform} & \textbf{\# posts} & \textbf{\# users} & \textbf{\# cascades} &\textbf{Duration}\\
     \hline
     Trump2020, \#MAGA, Dems, \#LiberalismIsAMentalDisorder & Tweets discussing the build-up to the 2020 US Elections. The content contains controversies, misinformation, and clashes between the two extreme sides of the US political spectrum.\\\hline
     Gaza, \#WeLoveIsrael, \#WeStandWithIsrael & These comprise the discussions, support and opposing comments around the Gaza-Israel clashes in 2019, along with the political crisis in Israel. \\\hline
     \#metoo, Epstein & Discussions revolving around the famous \#meToo movement with people coming out against sexual harassment and hate spread around the topic across the world.\\\hline
     Brexit, \#EUElections2019, \#PeoplesVote & These keywords are part of the tweets about people's opinions on the Brexit referendum, the controversies around it and clashes between people on both sides of the Brexit debate.\\\hline
     \#prolife, \#prochoice & Opinions, controversies, and clashes between both spectrums of abortion are the major contributors to these hashtags on Twitter. Moreover, the introduction/amendments of abortion laws across various states of the US gave rise to a majority of the content of this topic.\\\hline
     area 51, \#StormArea51 & Discussions around the famous "StormArea51" American Facebook event that took place on Twitter, along with controversial conspiracies around the presence of aliens in the US Air Force facility in Nevada.\\\hline
    \end{tabular}
    \endgroup
    \caption{A brief overview of the hashtags/keywords in the Twitter dataset we use, along with a brief description of each group. Hashtags/keywords in the list are further grouped on the basis of the similarity of the topics they represent.}
    \label{tab:twit_hashtag_detail}
\end{table}

\subsection{Hate Scoring and Characterization}
For the majority of our analysis, characterizing the content in terms of the hatefulness is of utmost importance. The current section provides details on how we assign hatefulness score to posts and, finally, characterize posts/users into three degrees of hatefulness each.

\subsubsection{Posts}
\label{sec:post_hate_characterize}
We subject each post in each dataset to three state-of-the-art hate speech classification systems, namely Davidson's \cite{davidson2017automated}, Waseem's \cite{waseem2016hateful} and Founta's \cite{Fountana} systems. Each system, based on its paradigm, generates a confidence score for each post, which is used to decide whether that post is found to be hateful or not by that system.

Furthermore, we use these systems' classifications to characterize each post into three buckets of hatefulness -- \textbf{high:} if two or more systems found the post hateful, \textbf{medium:} if one and only one system found the post hateful, and  \textbf{non:} if none of the systems found the post hateful.

\subsubsection{Users}
In order to characterize users into three buckets of hatefulness (low, medium, and high), we extract, for each user, their posts and the hatefulness characteristics, as explained in Section \ref{sec:post_hate_characterize}. We then classify each hateful user (must have posted at least one hateful content within the duration of the dataset collection span) as follows -- \textbf{high:} if the user posted five or more hateful posts (medium and/or high), \textbf{medium:} if the user posted two or more hateful posts, and \textbf{low:} if the user posted only one or no hateful post, as suggested in \cite{10.1145/3292522.3326034}.

\textcolor{black}{{\subsection{Hate Scoring Annotation and Validation}
To validate our approach for automated hate speech classification, we sampled a subset of 500 posts from each of the social networks. Three annotators were employed; all of them were in an age-group of 25-30, regular users of these platforms, and served the role of annotators for online toxicity detection previously. Each post was given a score of either 0 (non-hateful), 1 (medium hateful) or 2 (highly hateful) by each of these annotators. For each post, we then took the aggregate of the scores received and round off to the nearest integer to obtain the final annotation score. 
For annotation we set the following guidelines:
1. We classify abusive or derogatory posts targetting a community, gender, race, religion as highly hateful.
2. Posts that do not fall in above criteria, and express a persons opinion, inform about news, and are not offesnive to any person or community on any rights can be classified as non-hateful.
3. Posts that do not fall in above categories. This does not limit to posts which contain abuses or slangs, but not targetting a community or person directly, such posts are classified as medium-hateful.
An inter-annotator agreement of $0.78$ Cohen's $\kappa$ was found. Finally, we evaluate our proposed method of hatefulness scoring using the manually annotated data. The F1 scores of our model foor different platforms are as follows: $0.70$ for Reddit, $0.65$ for Gab, and $0.72$ for Twitter.}} \textcolor{black}{With the {\em balanced accuracy metric} provided by {Scikit-learn}\footnote{https://scikit-learn.org/stable/modules/generated/sklearn.metrics.balanced\_accuracy\_score.html}to handle label imbalances we get the following scores: $0.75$ for Reddit, $0.70$ for Twitter, and $0.72$ for Gab.}

Multiple previous studies have pointed toward the fact that hate speech classifiers trained using a specific training dataset annotated to identify specific types of hate speech fail to generalize when the data distribution changes due to shifting in target or time~\cite{zhang2019hate, florio2020time}. The very definition of hate speech is highly sensitive to multiple factors: considered target of hate (racism, sexism, anti-semitism, etc.), time-frame of the data (different events at different times instantiate different types of hateful discourse), type of content (forum post vs. microblogging vs. long articles) and many more. For example, based on a specific event, hatemongers might come up with disrespectful name-calling terms for their targets. These terms fade out and give space to newer terms based on the ever-happenning world of the online platforms. With such a rapid distribution shift, it is hard to achieve near-perfect classification performance using off-the-shelf classifiers.

\begin{figure}
\centering
\includegraphics[width=\textwidth]{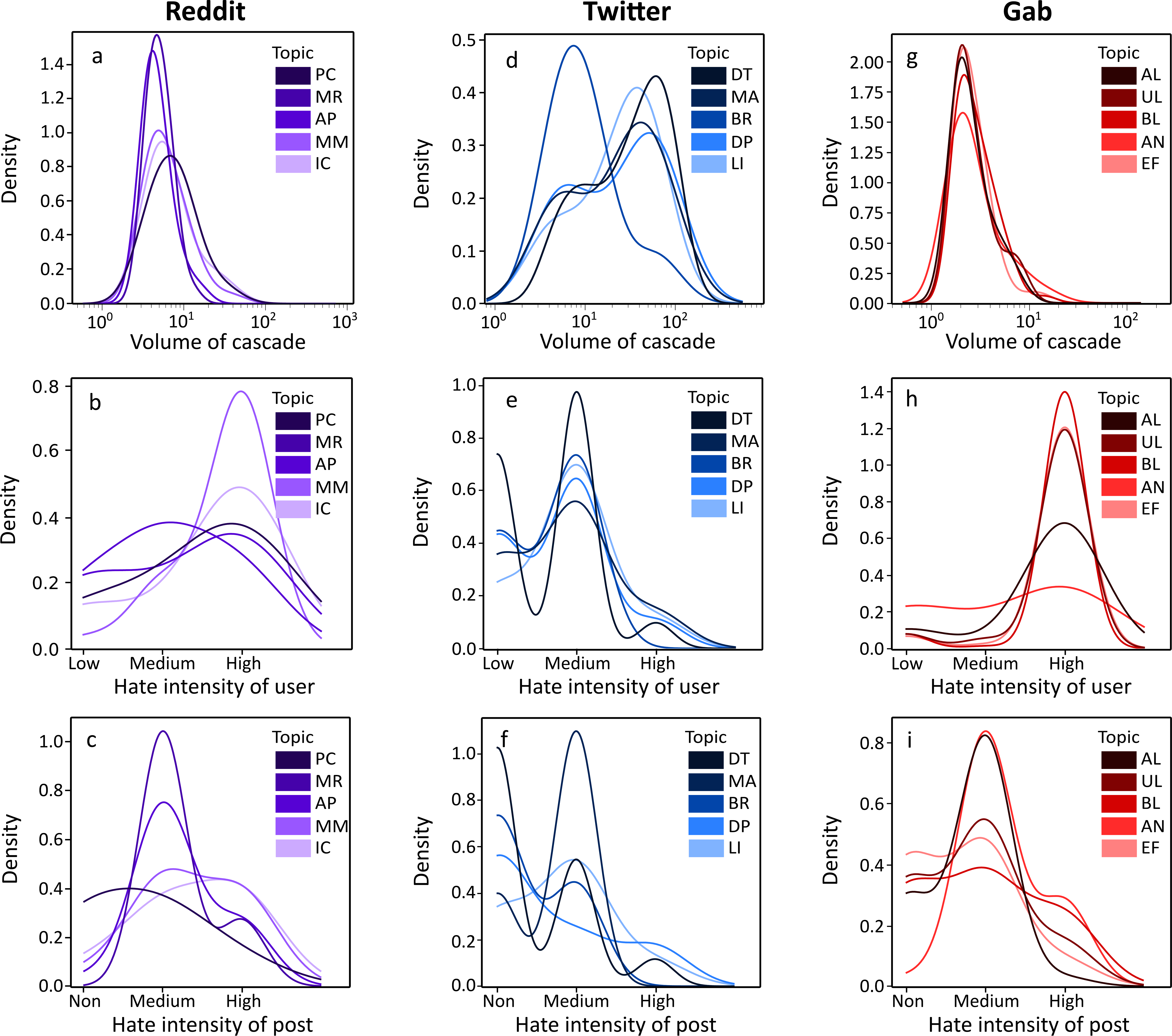}
\caption{{\color{black}Density distribution of hate intensity of source posts, source users as well as cascade volumes for top occuring topics in each platform. {\bf a., d., g.} Cascade volume distribution for the three social network for the selected topics. {\bf b., e., h.} Distribution of hate intensity of user for the three social netwokss across the selected topics. {\bf c., f., i.} Distribution of hate intensity of source post for the three social networks across the selected topics. We cover five topics for each social network -- Reddit (PC for pro-choice, MR for mens rights, AP for anti-abortion and pro-life, MM for mens mental health, IC for Illuminati conspiracy theories), Twitter (DT for Donald Trump, MA for MAGA, BR for Brexit, DP for Democratic Party, LI for \#Liberal), and Gab (AL for anti-abortion laws, UL for US border laws, BL for black lives matter, AN for antisemitism, EF for election fraud).}}
\label{fig:topic_analysis_response}
\end{figure}

\textcolor{black}{\subsection{Distribution through Topical Analysis}
\label{sec:TopicalAnalysisResponse}
In Figure \ref{fig:topic_analysis_response}, we study the density distribution of degree of hatefulness of source user, source post and the volume of the cascades for some of the top-occuring topics across the three social networks -- Reddit, Twitter and Gab. We cover topics ranging from politics to conspiracy theories, and black rights to antisemitism. A close look at these plots reveals that across majority of topics in Gab and Reddit, the user hatefulness density distribution peaks around the highly-hateful users. On the other hand, the same is true for medium-hateful users in Twitter. We observe spikes across both Gab and Twitter for high-hateful users for topics pertaining to the US politics; MAGA, Donald Trump for Twitter and Border laws and anti-abortion laws for Gab. For topics related to anti-abortion or pro-life, we notice that the distribution obtains a peak in Gab but the same is not observed for Reddit. This observation points towards the political inclination of users posting on the respective platforms, and the content they like to engage with. The general trend for the high-hateful users driving the spread of information still persists when observing the distribution of cascade volumes for social networks. However, the magnitude of cascade volumes is intrinsic to the network under consideration.
}

\section{Echo Chamber Detection}
\label{sec:topic_detection_si}
We propose a novel method for the detection of echo chambers in a social network. Given that we have access to the interaction network and the content shared between the users, the main idea around which we build our method is to automatically detect interactions based on similar topics and further extract groups of highly-clustered users in the network that take part in said interactions.

Let $U$ be the set of all users in a network, $C$ be the set of all content written on that network, and $C^u \subseteq C$ be the set of all content (posts or comments) written by  user $u \in U$ on that network. Finally, let $EC$ be the set of all encoded real-valued vectors, and $T$ be the set of topics as generated using $EC$, where $\forall t \in T, t \subseteq EC$.

\subsection{Topic Detection}
We pass each $c \in C$ through a pretrained natural language encoder to convert each piece of content to a uniquely-encoded real-valued vector $e \in EC_{NLE}$ of size 512 \cite{use}. Further, we apply principal component analysis (PCA) and uniform manifold approximation and projection (UMAP) over each vector, to reduce it to a smaller vector $e \in EC$ of size 64.

Let $NLE(\cdot)$ denote the natural language encoder that we use (Universal Sentence Encoder \cite{cer2018universal}, in our case), $PCA(\cdot)$ and $UMAP(\cdot)$ denote the functions for reducing the encoded vectors to a smaller size \cite{umap}.

\begin{equation}
    EC_{NLE} = \bigcup_{c \in C} NLE(c)
\end{equation}
\begin{equation}
    EC = \bigcup_{e \in EC_{NLE}} PCA(UMAP(e))
\end{equation}

We represent all of the textual content of the network in the form of these reduced vectors. 
We then perform clustering (HDBSCAN \cite{mcinnes2017hdbscan}, in our case) on the set $EC$ to find groups of similar content in terms of context and containing similar terms/phrases as detected by the natural language encoder. 

Let $CLS(\cdot)$ denote the clustering algorithm we use, which produces groups of the encoded vectors.

\begin{equation}
    T = CLS(EC)
\end{equation}

Each group represents a collection of posts corresponding to a topic discussed by the users, which we use to further extend to clusters of users discussing common topics. \textcolor{black}{To validate the quality of topical clusters created with the method above, we randomly sample some topics across the three social networks and analyse the content that is classified under them. We observe that the content clustered under a topic is similar in nature. Moreover, not only does the content refer to a similar event in time but also share the same ideology. We have uploaded multiple examples from each sample topic at \href{https://github.com/LCS2-IIITD/Hate-mongerer-and-echo-chambers/tree/main/Sample of Topic Clusters}{\underline{https://github.com/LCS2-IIITD/Hate-mongerer-and-echo-chambers/tree/main/Sample of Topic Clusters}}, within the codebase. The files are named with topics identified by the authors, using manual inference of content.}

\subsection{Extending Topics to Clusters of Users}
In the existing literature, an echo chamber is defined as a group of users who share the same opinion and reinforce their own beliefs \cite{Choi2020}. We cluster groups of users who share content on multiple topics multiple times. We argue that with this approach, we can segment users in both the criteria of echo chamber detection, i.e., ideological homophily and selective exposure. Since users share content with their own intent, they are being selective to the kind of content they want to react to. And since the topics were clustered using semantic information, the posts in a topic share the same ideology. The users' groups that are initially constructed in this manner are identified as candidate echo chambers. One issue we face is that many users are common across multiple topics, and some of the clusters identified have over 90\% similarities in terms of mutual users. We come up with a simple heuristic to reduce the number of unique candidate echo chambers and combine multiple echo chambers if they share a commonality in terms of users or topics above a specified threshold. For our experiments we combine two candidate echo chambers if the Jaccard Coefficient for the set of users belonging to the candidate echo chambers is greater than a threshold (0.7 in our case). The remaining clusters obtained after the reduction are finally classified as echo chambers.

\subsection{ Echo Chamber Network}
\begin{figure}
\centering
\includegraphics[width=\textwidth]{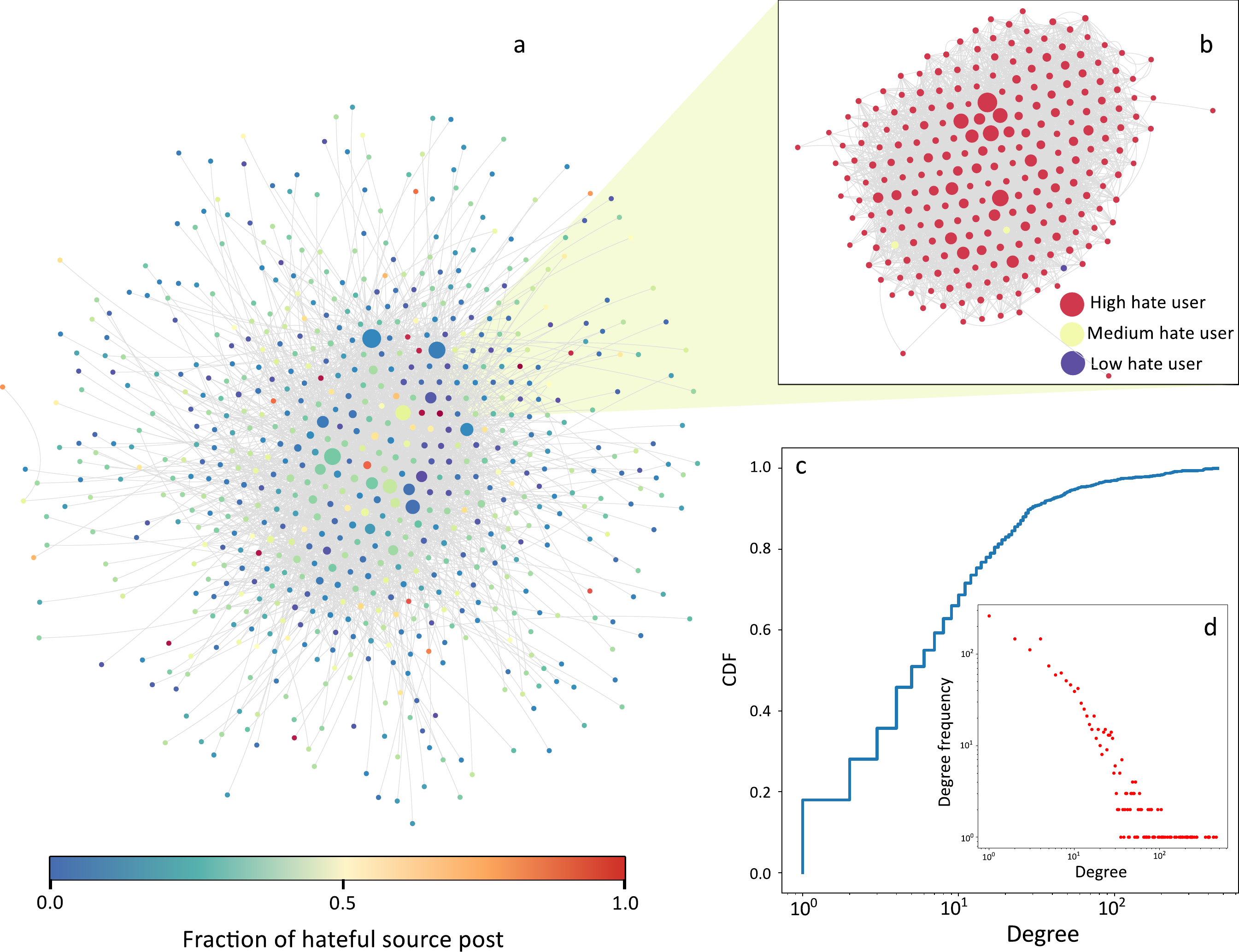}
\caption{Distribution of hate among echo chambers in Gab. {\bf a.} A sample network of echo chambers; each node represents an echo chamber color coded with the fraction of hateful source posts posted by members of the echo chamber; an edge between two echo chambers denotes common users. {\bf b.} A user interaction network within an example echo chamber; each node being a user with edges defined by {\em reply-to} interactions. {\bf c.} and {\bf d.} show the degree distributions of the network shown in {\bf a.}}
\label{fig:gab_network}
\end{figure}

\begin{figure}
\centering
\includegraphics[width=\textwidth]{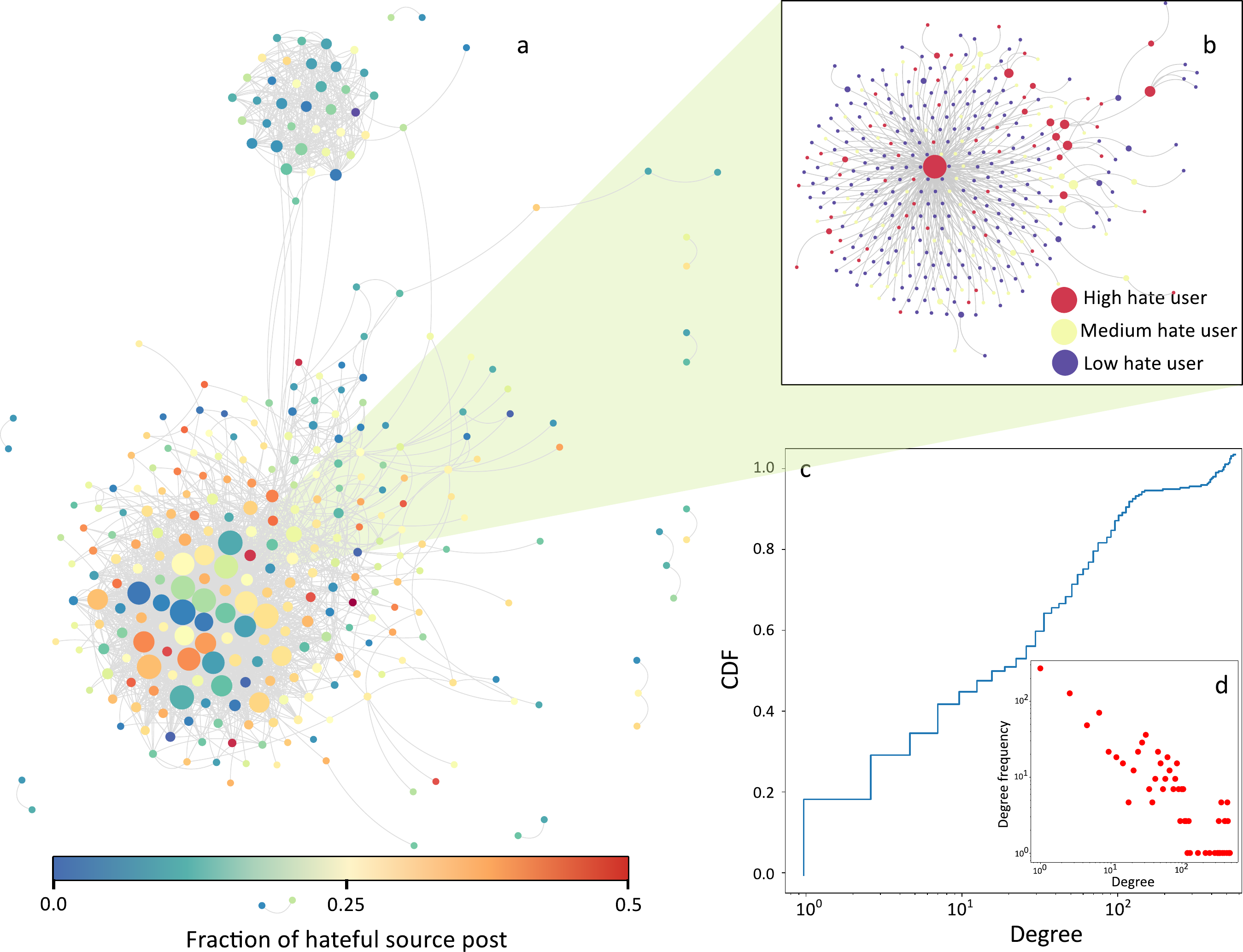}
\caption{Distribution of hate among echo chambers in Twitter. {\bf a.} A sample network of echo chambers; each node represents an echo chamber color coded with the fraction of hateful source posts posted by members of the echo chamber; an edge between two echo chambers denotes common users. {\bf b.} A user interaction network within an example echo chamber; each node being a user with edges defined by {\em reply-to} interactions. {\bf c.} and {\bf d.} show the degree distributions of the network shown in {\bf a.}}
\label{fig:twitter_network}
\end{figure}

We capture the relations between echo chambers in the form of an echo chamber network. We model this network in the form of an undirected weighted graph $G = (V, E, W)$, where $V$ is the set of echo chambers, $E$ is the set of edges, where $e_{ij} \in E$ denotes the presence of common users between $V_i$ and $V_j$. $W_{ij}$ denotes the weight of an edge $e_{ij}$, which is the number of common users between $V_i$ and $V_j$. Here, we only connect disjoint rumors, i.e., if $\exists e_{ij}$, then $T(V_i) \cap T(V_j) = \emptyset$, where $T(X)$ denotes topics composing echo chamber $X$. We notice that the networks created are very dense (for Gab and Reddit). To obtain a better visualization, we use a backbone extraction method \cite{doi:10.1073/pnas.0808904106} to get the important links (see Figures \ref{fig:gab_network} and \ref{fig:twitter_network}).

\section{K-core Decomposition}
\label{sec:k-core_decomposition}
A subgraph is said to be $k$-core or a core of order $k$ if and only if all the vertices of the subgraph have a minimum degree of $k$, and it is the largest possible subgraph satisfying that condition. 

Mathematically, we can define it as follows. Consider a graph $G = (V,E)$, where $V$ is the set of nodes, and $E$ is the set of edges connecting these nodes. Consider a subgraph $H = (A, E|A)$, where $A \subseteq V$. $H$ is a $k$-core of $G$ iff
$\forall v \in A$: $degree_H(v) \ge k$, and $H$ is the maximum subgraph satisfying the condition. 

$K$-core decomposition is a method in which we partition the graph into multiple cores by varying $k$. The corresponding cores are nested, i.e., $\forall i < j \Longrightarrow H_i \subseteq H_j$. It is not necessary for the subgraph to be connected in a core. The method helps us in extracting more central nodes. The higher the $k$-core number of a node, the more densely it is connected in the network.

\section{Significance Testing}
\label{sec:significance-testing}
\subsection{Volume of Cascades}
We hypothesize that the volume of cascades is impacted by the degree of hatefulness of a user and not impacted by the degree of hatefulness of the post. We use the Kolmogorov-Smirnov test to check whether our hypothesis is statistically significant.

To validate the impact of the degree of hatefulness of source users on cascade volume, we consider the continuous distribution of cascades from hateful source users ($F(x)$) and continuous distribution of cascades from non-hateful source users ($G(x)$). We define the null hypothesis as if the two distributions are identical, i.e., $F(x) = G(x)$, and the alternate hypothesis as $F(x) \neq G(x)$.

Similarly, to validate the impact of the degree of hatefulness of source posts on cascade volume, we consider continuous distribution of cascades from hateful source posts ($F(x)$) and continuous distribution of cascades from non-hateful source posts ($G(x)$). The null and alternate hypotheses are defined in the same manner to validate the impact of the degree of hatefulness of a source user.

From our analyses, we  conclude that the impact of the degree of hatefulness of a source user on volume of the cascades is statistically significant for all three social networks, with each reporting $p$-value < $0.02$. Hence, we can clearly reject the null hypothesis in this scenario. In contrast, for the degree of hatefulness of source posts, we get $p$-values > $0.05$ for Gab and Twitter, making us unable to reject the null hypothesis.

We conduct similar experiments for cascade width and height for all social networks. We find that impact of the degree of the hatefulness of a source user on cascade width is statistically significant for all three social networks with $p$-values < $0.05$. For the impact of the degree of the hatefulness of a source post on cascade width, we get $p$-values > $0.05$ for Gab and Twitter, which is synonymous with the scenario for cascade volume.
Next, we analyze the impact of the degree of the hatefulness of a source user on cascade height; we get $p$-values < $0.01$ for both Reddit and Gab, which rejects the null hypothesis. Regarding the hatefulness of a source post on cascade height, we get $p$-value > $0.05$ for Gab, indicating that the null hypothesis is accepted.

\subsection{Volume of Cascades of Echo Chamber Users}
We hypothesize that the cascade volume distributions from all highly-hateful source users and those source users who belong to echo chambers are different. We use the Kolmogorov-Smirnov test to check whether our hypothesis is statistically significant.

To validate the impact when the source user of a hateful post belongs to an echo chamber, we consider the continuous distribution of cascades from all hateful source users ($F(x)$) and the continuous distribution of cascades from hateful source users that belong to an echo chamber as ($G(x)$). We define the null hypothesis that the distribution of cascades of source users belonging to echo chambers is similar to the distirbution of cascades of hateful users, i.e., $F(x) = G(x)$, and the alternate hypothesis as $F(x) \neq G(x)$. 

So, in the case, as our null hypothesis isn't rejected, we can say that the distribution of volumes of cascades with highly-hateful source users that belong echo chambers is similar to the distribution of volumes of cascades from all highly-hateful source users, which is the case that we observe, as we get $p$-value > $0.05$.

In contrast, when we conduct the same experiment by replacing source users belonging to echo chambers with source users not belonging to echo chambers, we observe that we get $p$-value  < $0.05$, which rejects the null hypothesis, and we accept the alternate hypothesis. 

Hence, we can conclude that when highly-hateful source users belonging to echo chambers post content on social networks, the cascade formed are more similar to the cascades formed by posts from highly-hateful users.

\subsection{Impact of Degree of Hatefulness of a Source User on Fraction of Hateful Interactions when the Source Post is Hateful}
\begin{figure}
\centering
\includegraphics[width=\textwidth]{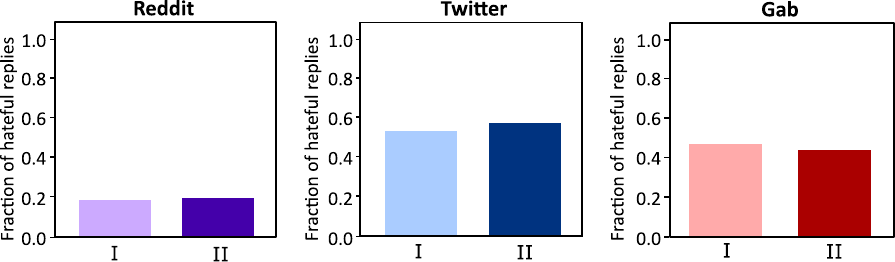}
\caption{Fraction of hateful interactions within cascades generated by hateful posts of
(I) a low-hate user vs. (II) a high-hate user.}
\label{SI:fraction_hateful_reply}
\end{figure}

%Through the above analysis, we have shown that the degree of the hatefulness of the source user significantly impacts the volume of cascades. Furthermore, 
We hypothesize that the degree of hatefulness of a user also impacts the fraction of hateful interactions that occur on a hateful post, as shown in Figure \ref{SI:fraction_hateful_reply}. To measure if the effect is significant, we run a Kolmogorov-Smirnov test to validate our hypothesis.

To validate the impact of the degree of the hatefulness of a user on the fraction of hateful interactions, we calculate the size of each cascade, the degree of the hatefulness of the source user, the degree of the hatefulness of the source post, and the fraction of hateful interactions in the cascade. We consider the distribution of the fraction of hateful interactions for only a high-hateful post from a high-hateful source user as $F(x)$, and the distribution of the fraction of hateful interactions for a high-hateful post from a low-hateful source user as $G(x)$. We define the null hypothesis as the degree of the hatefulness of a source user having no impact on the fraction of hateful interactions, i.e. $F(x) = G(x)$, and consequently, the alternate hypothesis being $F(x) \neq G(x)$. However, we observe that we get a KS statistic of 0.869 for Reddit, 0.878 for Twitter, and 0.948 for Gab, all with $p$-values < $0.001$. Hence, we can reject the null hypothesis.

\subsection{Core-wise distribution of user hate intensity}
To validate the correlation between core number and hatefulness of a user, we calculate the Spearman correlation coefficient between the distribution of core numbers and degree of hatefulness of a user. We get Spearman $rho$ values as follows: 0.68 in Reddit, 0.30 in Twitter, and 0.77 in Gab, all with $p$-values < $0.001$. The values for Spearman $p$, vary between $-1$ and $1$, with $0$ indicating no correlation. Despite the correlation being low for Twitter, we still get high correlation for both Reddit and Gab. We can say that both Reddit and Gab show a monotonic increase in highly-hateful users as the core number increases.

\textcolor{black}{\subsection{User Metadata Analysis}
Several studies have established relationships between cascade growth and different attributes of the root user.  To establish the validity of user hatefulness as a viable feature of cascade growth, we seek to measure how much information it shares with other cascade predictors. Follower count and age of the user account are two prominent ones among such attributes~\cite{age-and-follower, follower}. We compute Normalized Mutual Information (NMI) between a pair of variables; a near zero NMI would suggest independent distributions.
% showed that the volume of the cascade generated by a post is positively impacted by the number of followers the user has. Through our work we have shown that the hatefulness of the source user also impacts the volume of the cascades generated by their post. To validate that properties like follower count and similar user metadata do not impact the hatefulness of a user, we calculate the mutual information between these metadata properties and the hatefulness of a user. Note that such information is not applicable for users of Reddit; therefore, for this analysis, we only consider Gab and Twitter. To standardize the experiments further, we present our analysis for two commonly-used metadata properties -- follower count and age of the account. 
Hatefulness of the user shares a very low NMI with the follower count: $0.034$ for Gab and $0.044$ for Twitter. Similar patterns are observed in case of user account age as well: $0.045$ and $0.055$ NMI with user hatefulness in case of Gab and Twitter, respectively. We do not elect the following count of the accounts, as accounts of popular users like celebrities, sports personalities and politicians tend to have lower following counts which is also the scenario with people less active on social media.} 
% Furthermore, recent studies \cite{10.3389/fphy.2022.951729} also stated that follower count might not be an accurate parameter in predicting cascades. To this extent, we believe that inclusion of hatefulness as a predicting parameter for cascade growth could be of value in future work.}\todo{Need para needs polishing}

%%% Each figure should be on its own page
\clearpage

\clearpage
% \begin{figure}
% \centering
% \includegraphics[width=\textwidth]{HateDistributionPlotsForSupplementaryBasedOnSourceUserAndSourcePost.pdf}
% \caption{Fraction of type of interaction (hateful or non-hateful) characterized by hatefulness of source post (subplots a., c., e.) vs hatefulness of source user (subplots b., d., f.). 
% For each of the three networks, hateful sources (be it posts or users) are more likely to attract a higher fraction of hateful interactions.}
% \label{SI:fraction_of_hateful_interactions}
% \end{figure}

\clearpage

\clearpage

\clearpage

\clearpage

\clearpage

% \begin{figure}
% \centering
% \includegraphics[width=\textwidth]{Echo_hate_width_density.pdf}
% \caption{For each network, {\bf a.}, {\bf c.}, and {\bf e.} separate out the density distribution of cascade width for highly-hateful source users when they do or do not belong to any echo chamber. {\bf b.}, {\bf d.}, and {\bf f.} compare the fraction of interactions in the cascades originated by highly-hateful echo chamber members coming from other members vs. non-members; here, in Gab and Twitter, echo chambered users are more likely to participate in cascades originated by fellow echo chamber members.}
% \label{fig:cascade_width_echo}
% \end{figure}

% \begin{figure}
% \centering
% \includegraphics[width=0.67\textwidth]{Echo_hate_height_density.pdf}
% \caption{For each network, {\bf a.} and {\bf c.} separate out the density distribution of cascade height for highly-hateful source users when they do or do not belong to any echo chamber. {\bf b.} and {\bf d.} compare the fraction of interactions in the cascades originated by highly-hateful echo chamber members coming from other members vs. non-members.}
% \label{fig:cascade_height_echo}
% \end{figure}

\clearpage
\clearpage
\clearpage

% \begin{figure}
% \centering
% \includegraphics[width=0.67\textwidth]{height2width.pdf}
% \caption{}
% \label{fig:Height2Width}
% \end{figure}

% \begin{table}\centering
% \caption{This is a table}

% \begin{tabular}{lrrr}
% Species & CBS & CV & G3 \\
% \midrule
% 1. Acetaldehyde & 0.0 & 0.0 & 0.0 \\
% 2. Vinyl alcohol & 9.1 & 9.6 & 13.5 \\
% 3. Hydroxyethylidene & 50.8 & 51.2 & 54.0\\
% \bottomrule
% \end{tabular}
% \end{table}

%%% Add this line AFTER all your figures and tables

% \movie{Type legend for the movie here.}

% \movie{Type legend for the other movie here. Adding longer text to show what happens, to decide on alignment and/or indentations.}

% \movie{A third movie, just for kicks.}

% \dataset{dataset_one.txt}{Type or paste legend here.}

% \dataset{dataset_two.txt}{Type or paste legend here. Adding longer text to show what happens, to decide on alignment and/or indentations for multi-line or paragraph captions.}

\twocolumn
\bibliographystyle{acm}
\bibliography{main}
\end{document}